\documentclass[journal,onecolumn,11pt]{IEEEtran}

\pdfoutput=1
\usepackage{cite}
\usepackage{array}
\usepackage{url}
\usepackage{longtable}
\usepackage{multirow}
\usepackage{amsmath}
\usepackage{amssymb}
\usepackage{blkarray}
\usepackage{fancyhdr}
\usepackage{caption}
\usepackage{subcaption}
\usepackage{textcomp}

\usepackage{stfloats}

\usepackage{multirow}

\usepackage{verbatim}

\usepackage[pdftex]{graphicx}
\usepackage{setspace}

\usepackage{slashbox}

\usepackage[pdftex]{graphicx}
\usepackage{setspace}

\providecommand{\abs}[1]{\lvert#1\rvert}
\providecommand{\norm}[1]{\lVert#1\rVert}

\makeatletter
\def\blfootnote{\xdef\@thefnmark{}\@footnotetext}
\makeatother
\begin{document}


\title{Formulation and Steady-state Analysis of LMS Adaptive Networks for Distributed Estimation in the Presence of Transmission Errors}

\author{\IEEEauthorblockN{Saeed Ghazanfari-Rad and Fabrice Labeau}}

\maketitle

\begin{abstract}

This article presents the formulation and steady-state analysis of the distributed estimation algorithms based on the diffusion cooperation scheme in the presence of errors due to the unreliable data transfer among nodes. In particular, we highlight the impact of transmission errors on the least-mean squares (LMS) adaptive networks. We develop the closed-form expressions of the steady-state mean-square deviation (MSD) which is helpful to assess the effects of the imperfect information flow on on the behavior of the diffusion LMS algorithm in terms of the steady-state error. The model is then validated by performing Monte Carlo simulations. It is shown that local and global MSD curves are not necessarily monotonic increasing functions of the error probability. We also assess sufficient conditions that ensure mean and mean-square stability of diffusion LMS strategies in the presence of transmission errors. Moreover, issues such as scalability in the sense of network size and regressor size, spatially correlated observations, as well as the effect of the distribution of the noise variance are studied.

While the proposed theoretical framework is general in the sense that it is not confined to a particular source of error during information diffusion, for practical reasons we additionally study a specific scenario where errors occur at the medium access control (MAC) level. We develop a model to quantify the MAC-level transmission errors according to the network topology and system parameters for a set of nodes employing a backoff procedure to access the channel. To overcome the problem of unreliable data exchange, we propose an enhanced combining rule that can be deployed in order to improve the performance of diffusion estimation algorithms by using the knowledge of the properties of the transmission errors.  


\end{abstract}


\let\thefootnote\relax\footnote{The authors are with the Department of Electrical and Computer Engineering, McGill University, Montr\'eal, QC H3A 0E9, Canada. E-mail: \{saeed.ghazanfarirad@mail.mcgill.ca; fabrice.labeau@mcgill.ca\}.\\
\indent A preliminary version of this work considering a two-node network has appeared as a conference paper in the \emph{Proceedings of Asilomar Conference on Signals, Systems and Computers, Pacific Grove, CA, November 2012.}\\
\indent This work was supported by Hydro-Quebec, the Natural Sciences and Engineering Research Council of Canada and McGill University in the framework of the NSERC/Hydro-Quebec/McGill Industrial Research Chair in Interactive Information Infrastructure for the Power Grid.
}

\section{Introduction}
Deployment of wireless sensor networks (WSNs) as monitoring and diagnostic systems is receiving significant attention in recent years because of their clear advantage of being cost-effective and easy to deploy compared to the traditional wired-based ones. Furthermore, through the implementation of WSNs operating in a collaborative mode, a wide variety of applications such as smart grid, precision agriculture, intelligent transportation systems, disaster relief management, radar and target detection and tracking, etc, would benefit from highly reliable and flexible monitoring and diagnostic system that rapidly respond to the changing conditions as well as instantaneous faults.  Thus, distributed implementation of estimation algorithms are recently forming an active area of research in the context of distributed adaptive filtering. In the conventional estimation systems, the nodes collect measurements and send them to a fusion center for final centralized processing. The central node would also broadcast the information to the individual nodes. Hence, the sensors achieve an estimate that is as accurate as the one that would be obtained if each sensor had access to all the information available across the network. However, such a traditional scheme has the disadvantages of a high communication overhead and power consumption. Furthermore, the centralized implementation is not scalable in terms of the communication bandwidth and computational complexity and lacks robustness in terms of the link failures. 

A different approach for information exchange is the distributed in-network processing algorithms. In distributed processing, each node collects noisy observations related to a certain parameter or phenomenon of interest. The nodes would then communicate with their neighbors rather than a fusion center in order to arrive at an estimate of the parameter of interest. Distributed signal processing leads to significant saving in terms of bandwidth and power resources by reducing the communication overhead and the processing load \cite{li2002,culler2004,rabbat05}. Based on the topology of the network, different distributed algorithms can be implemented \cite{lopes_incr_2007,Lopes2008,Cattivelli2008,Cattivelli2010}. Throughout this paper, we study the diffusion mode of cooperation in which each node communicates with all its immediately adjacent neighbors according to the network topology \cite{xiao05,xiao06,Lopes2008,Cattivelli2008}.
Furthermore, we concentrate on the combine-then-adapt (CTA) diffusion algorithm. The CTA implementation consists of two steps: first, the local estimate and the ones obtained from the neighbors are linearly combined and in the second step the adaptation is performed. The CTA algorithm has been first proposed in \cite{sayed07,lopes07,lopes_ssp07,Lopes2008} and later its modified versions appeared in \cite{Cattivelli2010,catti08,catti_kal10}. 

There exists previous literature for performance analysis of diffusion algorithms, but all these studies consider an ideal and error-free transmissions \cite{Lopes2008,Cattivelli2008,Cattivelli2010}. Some studies have already considered the diffusion algorithms with noisy information exchanges \cite{khali12,zhao12noise}. Theses studies introduce an additive noise component to model the noisy link in different steps of the diffusion algorithms. The analytical framework that we propose differs from the literature in the way that we model the imperfect information exchange. In particular, we incorporate the discrete-event random errors into the formulation which enables the analysis to account for the random transmission errors arising from several factors such as collisions, node failure, link failure and link congestion.

When dealing with practical and error-prone transmissions, a careful modeling and performance analysis is required to highlight the effects of transmission errors on the behavior of the diffusion algorithms. Since adaptive filters are inherently nonlinear time-varying systems, often theoretical development of a single stand-alone LMS filter is a difficult task and involves a number of assumptions on the observed data. This mathematical hurdle would be more challenging when dealing with the coupling effects arising from the diffusion algorithm and the inherently time-varying transmission errors. The main contribution of our research is to formulate and analyze the performance of LMS adaptive networks for distributed estimation considering transmission errors. We already examined a two-node network with regressor vectors of size $M \times 1$ which can be considered as a special case of this article that focuses on a network including $N$ nodes \cite{saeed_asilomar2012}. We notice that ideal error-free scenarios as studied previously \cite{lopes_incr_2007,Lopes2008,Cattivelli2008,Cattivelli2010} can be considered as a special case of the formulation in this paper. Meanwhile the mathematical foundation proposed in this article brings the advantage of avoiding the procedure of inverting a matrix of size $N^2 M^2 \times N^2 M^2$ (as required in \cite{lopes_incr_2007,Lopes2008,Cattivelli2008,Cattivelli2010}) which might lead to computational problems for large $N$ and/or $M$. Furthermore, we do not impose the constraint of spatially independent regressors across the distributed nodes.

The remainder of the paper is organized as follows. In Section \ref{analysis}, we formulate the problem of diffusion algorithms over distributed adaptive networks in the presence of transmission errors. In this section, we provide the mean analysis, mean-square analysis and closed-form derivation of the steady-state MSD. We also derive sufficient conditions to ensure the stability of the algorithm in the mean and mean-square sense. In Section \ref{Modeling transmission errors}, We quantify transmission errors by studying a case scenario where errors occur at the MAC layer. In Section \ref{Performance improvement}, we aim at improving the performance of diffusion estimation algorithms by introducing a combining policy that accounts for transmission errors. In order to verify the accuracy of the proposed theoretical framework and observe the impact of transmission errors on the distributed estimation algorithms, we present the simulation results in Section \ref{sim}. Finally, we conclude the paper in Section \ref{con}.

\section{Analysis of diffusion LMS over a Network including $N$ nodes}
\label{analysis}

\subsection{Problem Formulation}
\label{sec:prob_form}


Consider a distributed network with a set of nodes $\mathcal{N}=\{1,2,\ldots,N\}$ and a predefined topology including $L$ links. In a typical wireless deployment, a link $(k,l), k,l \in \mathcal{N}$ exists between two nodes if and only if the physical distance between the end nodes is less than the maximum radio transmission range.  Let $\mathcal{L}=\{(k,l)\}, k,l \in \mathcal{N}$  denote the set of all links of the network. Let $\mathcal{N}_k$ denote the set of nodes in the neighborhood of node $k$ (i.e., those with which node $k$ has a link) including node $k$ itself. The objective of the network is to estimate the unknown parameter vector $\boldsymbol{w}^o$ in a distributed manner from measurements of $N$ nodes. In a diffusion strategy every node $k$ at each time $i \ge 0$ employs some mixing coefficients to combine estimates from its neighborhood $\mathcal{N}_k$ \cite{Lopes2008}. However, in practice not all of the linked nodes are able to send their estimates to node $k$ due to the errors at different layers of the communication protocol stack. Consequently, unlike the ideal situation in which no error occurs, each iteration of the adaptive algorithm builds up a different set of mixing coefficients that depends on the error probabilities. 

Let $\mathcal{S}_{k,i} \subseteq \mathcal{N}$ denote the set of neighbors of node $k$ that successfully transmit their information to node $k$ at time $i$ including node $k$. We introduce the adaptive filter of node $k$ in a distributed network based on diffusion LMS with transmission errors as follows: 
\begin{equation}
\label{nodek}
\boldsymbol{w}_{k,i+1}=\boldsymbol{\phi}_{k,i}+\mu_{k} e_{k,i} \boldsymbol{u}_{k,i}, \quad i \ge 0
\end{equation} where
\begin{align}
\label{nodekphi}
\boldsymbol{\phi}_{k,i}&=\sum \limits _{l \in \mathcal{S}_{k,i}} {a_{k,l,i} \boldsymbol{w}_{l,i}}\\
\label{nodeke}
e_{k,i}&=d_{k,i}-\boldsymbol{\phi}_{k,i}^{T}\boldsymbol{u}_{k,i}\\
\label{nodekd}
d_{k,i}&=v_{k,i}+\boldsymbol{w}^{o^T}\boldsymbol{u}_{k,i}
\end{align} where $\boldsymbol{w}_{k,i},~k=1,2,\ldots,N,~i \ge 0$ are the $M$-dimensional coefficient weight vectors of the adaptive algorithm at node $k$ and iteration $i$, $\mu_{k}$ is the step size at node $k$, $a_{k,l,i},~l=1,2,\ldots,N$ are the mixing coefficients at iteration $i$, $\boldsymbol{\phi}_{k,i}$ are the intermediate variables to obtain the new weight vectors after information exchange among nodes, $\boldsymbol{u}_{k,i}$ are the $M$-dimensional input vectors, $e_{k,i}$ are the error signals, $v_{k,i}$ are the noise signals and $d_{k,i}$ are the desired signals obtained from the unknown weight vector $\boldsymbol{w}^{o}$ through the linear regression model (\ref{nodekd}). The superscript $(.)^T$ denotes transpose operation.

Throughout this article we assume that a failure occurs in information flow from node $l$ to node $k$ at iteration $i$ with probability $p_{k,l}$. The difficulties that cause such failures or transmission errors may include, but are not limited to: mobility of nodes and time-varying network topology, interference and multipath fading, signal attenuation at the physical (PHY) layer, packet loss at the MAC layer and attacks originated by attackers or intruders. Thus, it is reasonable to assume that in general such errors are independent non-identically distributed. However, our analysis is not restricted to independent errors assumption. Furthermore it is reasonable to assume that transmission errors that occur with probabilities $p_{k,l}$ and the measurement errors $v_{k,i}$ are independent. As an example, a node may observe a low measurement noise variance (for example if it is close to a target in tracking applications) but still experience high transmission error due to the high density of nodes in that area. In order to give insights to how to quantify transmission errors, we provide a model in Section \ref{Performance improvement} to assess $p_{k,l}$ when concentrating on the MAC-level errors.

We notice that unlike the previous formulation (ideal transmission) \cite{Lopes2008,Cattivelli2008,Cattivelli2010}, here, the intermediate variable $\boldsymbol{\phi}_{k,i}$ is constructed by a weighted sum over the set of nodes who have successfully transmitted their local information to node $k$. To further demonstrate the new formulation, we use a different way to express $\boldsymbol{\phi}_{k,i}$ as follows:
\begin{eqnarray}
\label{interpret}
\boldsymbol{\phi}_{k,i}=a_{k,k,i} \boldsymbol{w}_{k,i} +\sum \limits _{l \in \mathcal{N}_{k}\setminus\{k\}} {a_{k,l,i} \boldsymbol{w}_{l,i}}-\sum \limits _{l \in \mathcal{N}_{k}\setminus\{k\} } {\delta_{k,l,i} a_{k,l,i} \boldsymbol{w}_{l,i}}   
\end{eqnarray} where $\delta_{k,l,i}, k=1,2,\ldots,N, l\in \mathcal{N}_{k}\setminus\{k\} $ is a Bernoulli random variable with parameter $p_{k,l}$:

\begin{equation}
\label{modeltranserror}
\delta_{k,l,i}:=\begin{cases}
1 & \text{with probability} \quad p_{k,l}\\
0 & \text{with probability} \quad 1-p_{k,l}
\end{cases}
\end{equation}

The interpretation of the last term on the right hand side of (\ref{interpret}) is to eliminate the local weight vectors of those nodes that have not been able to successfully transmit their information to node $k$. The compact form of (\ref{interpret}) is the expression already stated in (\ref{nodekphi}) which builds $\boldsymbol{\phi}_{k,i}$ using a weighted sum of the local weight vectors over $\mathcal{S}_{k,i}$. In order to incorporate the transmission errors in the formulation of diffusion algorithm, we subsequently define some useful notation. Define $\Lambda=[\lambda_{kl}], k,l \in \mathcal{N}$ as the $N \times N$ adjacency matrix representing the network connectivity, i.e, each entry $\lambda_{kl}$ is 1 if there is a link between nodes $k$ and $l$ and is 0 if there is not. Assume that $E$ is the number of $1$'s not located on the main diagonal of $\Lambda$, i.e., $E=2L$. Motivated by the aforementioned discussion and assumptions regarding unreliable transmissions, we define a set of possible events $\mathcal{E}=\{e_1,\ldots,e_V\}$, their corresponding probabilities $\mathcal{P}=\{p_1,\ldots,p_V\}$, where $V=2^{E}$ and the set of combining matrices $\mathcal{A}=\{\mathcal{A}_{1},\ldots,\mathcal{A}_{v}\}$. We also introduce the set $\mathcal{V}=\{1,\ldots,V\}$ whose $j^{th}$ element corresponds to the occurrence of event $e_{j}, j=1,\ldots,V$. We note that $\mathcal{A}_{j}=[a_{k,l,j}],~j=1,\ldots,V$ is the $N \times N$ combining matrix that collects the nonnegative mixing coefficients of diffusion update followed by event $e_{j}$ during information exchange period satisfying
\begin{equation}
\label{constraintona}
\sum_{l\in\mathcal{N}_k}{a_{k,l,j}}=1,~k\in\mathcal{N},~\text{for all}~j.
\end{equation}
 
It follows from the above diffusion algorithm that each entry $a_{k,l,j}$ of matrix $\mathcal{A}_{j}$ represents the weight given to node $l$ in order to find the intermediate variable at node $k$ conditioned that event $e_{j}$ occurred during information sharing. As an example, suppose that $e_{1}$ represents the event in which all transmissions fail due to congested links. Under the independent errors assumption, the probability associated to this case is $p_{1}=\prod_{(k,l) \in \mathcal{L}} {p_{k,l}}$ and it follows that $\mathcal{A}_{1}=\mathcal{I}_N$, where $\mathcal{I}_N$ is the $N \times N$ identity matrix, i.e., for this iteration each node establishes the update only according to its local observation. 

Regarding the statistics of the measurement data and noise signals, we assume that the regressors $\boldsymbol{u}_{k,i}$ are temporally independent identically distributed (i.i.d.) zero-mean white Gaussian random variables with covariance matrices $\mathcal{R}_{u_k}=E[\boldsymbol{u}_k\boldsymbol{u}_{k}^{T}]=\sigma^{2}_{u_k}\mathcal{I}_{M}$. However, we explore the spatial correlation between nodes by assuming that two locally observed vectors $\boldsymbol{u}_{k}$ and $\boldsymbol{u}_{l}$ are correlated Gaussian random vectors with cross-correlation matrix $\mathcal{R}_{u_k,u_l}=\sigma_{u_{kl}}^2 \mathcal{I}_M,$ where $\sigma^2_{u_{kl}}=\rho_{kl} \sigma_{u_k} \sigma_{u_l}$ and $\rho_{kl}$ is the spatial correlation index between nodes $k$ and $l$. The noise signal $v_{k,i}$ is zero mean i.i.d. white Gaussian random variable with variance $\sigma^2_{v_k}$. The input vectors $\boldsymbol{u}_{k,i}$ and noise signals $v_{k,i}$ are temporally and spatially independent of each other.

Note that (\ref{nodekphi}) represents a linear combination of the received weight vectors to produce the intermediate variable $\boldsymbol{\phi}_{k,i}$ at node $k$ at iteration $i$. In general, the combiners may be nonlinear or even time-variant to reflect variations in network topology or to respond efficiently to nonstationary conditions \cite{Lopes2008}. In the following discussion the mixing coefficients are considered to be time-varying in order to capture the effects of transmission errors. We use the above formulation throughout the forthcoming sections to work out the detailed mean and mean-squared analyses of diffusion estimation algorithms in the presence of transmission errors.

\subsection{Mean Analysis}
\label{Mean analysis}
In this subsection, we provide the mean analysis which will be used later in order to find the expression for the steady-state mean-square deviation (MSD).  Using (\ref{nodek})-(\ref{nodekd}) and conditioned that transmission errors correspond to $e_{j}$, we can obtain a recursive expression to calculate $E[\boldsymbol{w}_{k,i}|e_{j}]$ as follows
\begin{equation}
\label{recursivewkcon}
E[\boldsymbol{w}_{k,i+1}|e_{j}]=a_{k,k,j}\rho_{k}E[\boldsymbol{w}_{k,i}]+\rho_{k} \sum_{\substack{  l \in \mathcal{N} \\ l \neq k  }  } {a_{k,l,j} E[\boldsymbol{w}_{l,i}]  }+\boldsymbol{c}_{k},
\end{equation} where
\begin{align}
\label{rhogeneral}
\rho_{k}&=1-\mu_{k} \sigma^{2}_{u_{k}},\quad k \in \mathcal{N}\\
\label{cgeneral}
\boldsymbol{c}_{k}&=\mu_{k} \sigma^{2}_{u_{k}} \boldsymbol{w}^{o},\quad k \in \mathcal{N}.
\end{align}

In order to find $E[\boldsymbol{w}_{k,i+1}]$, we consider all possibilities according to set $\mathcal{E}$ and replace~(\ref{recursivewkcon}) in the following equation
\begin{equation}
\label{mean_st_con}
E[\boldsymbol{w}_{k,i+1}]= \sum \limits_{j \in \mathcal{V}} {p_j E[\boldsymbol{w}_{k,i+1}|e_j]},
\end{equation} which yields
\begin{equation}
\label{recursivewk}
E[\boldsymbol{w}_{k,i+1}]=a_{k,k}\rho_{k}E[\boldsymbol{w}_{k,i}]+\rho_{k} \sum_{\substack{  l \in \mathcal{N} \\ l \neq k  }  } {a_{k,l} E[\boldsymbol{w}_{l,i}]  }+\boldsymbol{c}_{k},
\end{equation} where 
\begin{align}
\label{recursivewkcoeff1}
a_{k,k}=\sum \limits_{j \in \mathcal{V}} {p_j a_{k,k,j}},\quad a_{k,l}=\sum_{j \in \mathcal{V}} {p_j a_{k,l,j}},
\end{align} for all $k,l\in\mathcal{N}, l\neq k$. It follows from (\ref{constraintona}) that 
\begin{equation}
\label{sumis1}
\sum_{l \in \mathcal{N}_k} {a_{k,l}}=1,~k\in \mathcal{N}. 
\end{equation}
From Appendix~\ref{app_ma}, we conclude that

\begin{equation}
\label{ewss_f}
E[\boldsymbol{w}_{k,s}] = \boldsymbol{w}^{o},\quad k=1,2, \ldots, N,
\end{equation} 
i.e., the weights converge to the optimal value.


\subsection{Mean Stability}
\label{Mean Stability}
In the solution procedure of the previous subsection and in particular in using the Cramer's law in Appendix~\ref{app_ma}, we should verify that $z=1$ is not a root of the polynomial of order $N$ obtained from $det(\mathcal{E}_n)=0$. This polynomial can be written as $F(z)=\sum_{k=0}^N{f_k}z^{-k}$. Moreover, for stability in mean, it is required that all roots of $F(z)$ lie within the unit circle. Using the structure of $\mathcal{E}_n$ defined in (\ref{ensnew}), it is easily verified that $F(z)$ is the characteristic polynomial of $\mathcal{E}_{n,s}^\prime=[\rho_i a_{i,j}]_{N \times N},~i,j=1,2,\ldots,N$ and it is immediate that the roots of $F(z)$ are the eigenvalues of the square matrix $\mathcal{E}_{n,s}^\prime$. Let $\lambda_k,~ k=1,2,\ldots,N$ denote the eigenvalues of  $\mathcal{E}_{n,s}^\prime$. We also use $\rho(\mathcal{E}_{n,s}^\prime)$ to denote the spectral radius of $\mathcal{E}_{n,s}^\prime$. In the sequel, we find the sufficient condition that guarantees that the maximum absolute eigenvalue $\max_{1 \leq k \leq N} \abs{\lambda_k}$ or equivalently the spectral radius $\rho(\mathcal{E}_{n,s}^\prime)$ is less than one. This condition is sufficient to place the roots of $F(z)$ within the unit circle and hence ensure stability in mean. Considering (\ref{sumis1}), we notice that one interesting feature of the rows of $\mathcal{E}_{n,s}^\prime$ is that
\begin{align}
\label{meansumofm}
\sum_{l \in \mathcal{N}} {a_{k,l}}=\rho_k, \quad k=1,2,\ldots,N.
\end{align} Consider the induced infinity-norm of matrix  $\mathcal{E}_{n,s}^\prime$ defined as
\begin{equation}
\label{infinitynorm}
\norm{\mathcal{E}_{n,s}^\prime}_{\infty}=\max_{1\leq k \leq N} \sum_{l \in \mathcal{N}} {\abs{a_{k,l}}}.
\end{equation} It is also known from the characteristics of a matrix norm that
\begin{equation}
\label{speclessthannorm}
\rho(\mathcal{E}_{n,s}^\prime) \leq \norm{\mathcal{E}_{n,s}^\prime}.
\end{equation}

To satisfy $\max_{1 \leq k \leq N} \abs{\lambda_k} < 1$, we use (\ref{meansumofm})-(\ref{speclessthannorm}) to express the following condition:
\begin{equation}
\label{meancondition}
\abs{\rho_k} < 1 , \quad k=1,2,\ldots ,N,
\end{equation} which is equivalent to impose the following lower and upper bounds on the step-sizes
\begin{equation}
\label{meanconditionmu}
0 < \mu_k < \frac{2}{\sigma^2_{u_k}}, \quad k=1,2,\ldots ,N.
\end{equation}

Consequently, the important result can be stated as follows. In a network including $N$ nodes deploying distributed diffusion estimation algorithm (\ref{nodek})-(\ref{nodekd}) with combining weight matrices satisfying (\ref{constraintona}), in the presence of multiplicative transmission errors modeled as (\ref{modeltranserror}), the sufficient condition for mean stability is provided by (\ref{meanconditionmu}). It is important to note that (\ref{modeltranserror}) coincides with sufficient condition for mean stability when transmissions are assumed to be perfect \cite{Cattivelli2010}. In other words, transmission errors modeled as (\ref{modeltranserror}) do not diverge diffusion estimation algorithm (\ref{nodek})-(\ref{nodekd}) in the mean sense.

\subsection{Mean-Square Analysis}
\label{Mean-Square Analysis}
We aim at finding the closed form expressions for the steady-state MSD. Notice that the steady state MSD value for weight vector at node $k$ is defined as follows:
\begin{equation}
\label{msdvalues}
\text{MSD}_{k}=\lim_{i\to\infty}{E[(\boldsymbol{w}_{k,i}-\boldsymbol{w}^{o})^{T}(\boldsymbol{w}_{k,i}-\boldsymbol{w}^{o})]}.
\end{equation}
It is shown in Appendix~\ref{app_ew4uw} that we can write the following expression for $E[\boldsymbol{w}_{k,i+1}^{T}\boldsymbol{w}_{l,i+1}|e_{j}]$:
\begin{align}
\label{klgeneral_spatial_con}
E[\boldsymbol{w}_{k,i+1}^{T}\boldsymbol{w}_{l,i+1}|e_{j}]&=\eta_{kl}\{ \sum_{m \in \mathcal{N}} {a_{k,m,j} a_{l,m,j} E[\boldsymbol{w}_{m,i}^{T}\boldsymbol{w}_{m,i}]}+\sum_{m,n \in \mathcal{N}, m \neq n} { (a_{k,m,j} a_{l,n,j}+a_{k,n,j} a_{l,m,j}) E[\boldsymbol{w}_{m,i}^{T}\boldsymbol{w}_{n,i}] } \} \nonumber \\
&\quad +\sum_{m \in \mathcal{N}} { (a_{k,m,j}(\epsilon_l-\nu_{kl})+a_{l,m,j}(\epsilon_k-\nu_{kl}))  \boldsymbol{w^{o}}^{T}E[\boldsymbol{w}_{m,i}]  } + \nu_{kl} \boldsymbol{w^{o}}^{T}\boldsymbol{w^{o}},
\end{align} where \begin{align}	
\label{coeff_general_kl}
\eta_{kl}&=1-(\mu_k \sigma^{2}_{u_k}+\mu_l \sigma^{2}_{u_l})+ \mu_k \mu_l (\sigma^2_{u_k} \sigma^2_{u_l} + (M+1) \sigma^4_{u_{kl}}  ), \\
\label{coeff_general_klep}
\epsilon_k &=\mu_k \sigma^2_{u_k}, \\
\label{coeff_general_klnu}
\nu_{kl}&=\mu_k \mu_l (\sigma^2_{u_k} \sigma^2_{u_l} + (M+1) \sigma^4_{u_{kl}}).
\end{align}

In order to consider the set of all possible events during information exchange period, we write: \begin{equation}
\label{ew1w2colmain}
E[\boldsymbol{w}_{k,i+1}^{T}\boldsymbol{w}_{l,i+1}]= \sum_{j \in \mathcal{V}} p_j {E[\boldsymbol{w}_{k,i+1}^{T}\boldsymbol{w}_{l,i+1}|e_j]}.
\end{equation}

Using~(\ref{klgeneral_spatial_con}) and (\ref{ew1w2colmain}), it follows that:
\begin{align}
\label{klgeneral_spatial}
E[\boldsymbol{w}_{k,i+1}^{T}\boldsymbol{w}_{l,i+1}] & = \sum_{m \in \mathcal{N}} { c_{kl,mm} E[\boldsymbol{w}_{m,i}^{T}\boldsymbol{w}_{m,i}]} +\sum_{\substack{m,n \in \mathcal{N} \\ m \neq n}} { c_{kl,mn} E[\boldsymbol{w}_{m,i}^{T}\boldsymbol{w}_{n,i}] }  \nonumber \\
&\quad +\sum_{m \in \mathcal{N}} { c_{kl,om}  \boldsymbol{w^{o}}^{T}E[\boldsymbol{w}_{m,i}]  }+\nu_{kl} \boldsymbol{w^{o}}^{T}\boldsymbol{w^{o}},
\end{align}
where
\begin{align}	
\label{coeff_general_N_nodes}
c_{kl,mm}&=\eta_{kl} \sum_{j \in \mathcal{V}} {p_{j} a_{k,m,j} a_{l,m,j}},\\ 
\label{coeff_general_N_nodes_1}
c_{kl,mn}&=\eta_{kl} \sum_{j \in \mathcal{V}} { p_{j} (a_{k,m,j} a_{l,n,j} + a_{k,n,j} a_{l,m,j}) },\\
\label{coeff_general_N_nodes_klom}
c_{kl,om}&=\sum_{j \in \mathcal{V}} { p_{j} [a_{k,m,j}(\epsilon_l-\nu_{kl})+a_{l,m,j}(\epsilon_k-\nu_{kl})] },
\end{align} for all $k,l,m,n \in \mathcal{N},~k \neq l~\text{and}~m \neq n$. In a similar way, we can write:
\begin{align}
\label{kkgeneral_spatial}
E[\boldsymbol{w}_{k,i+1}^{T}\boldsymbol{w}_{k,i+1}] & = \sum_{m \in \mathcal{N}} {c_{kk,mm} E[\boldsymbol{w}_{m,i}^{T}\boldsymbol{w}_{m,i}]} +\sum_{\substack{m,n \in \mathcal{N} \\ m \neq n}} { c_{kk,mn} E[\boldsymbol{w}_{m,i}^{T}\boldsymbol{w}_{n,i}] }  \nonumber \\
& \quad + \sum_{m \in \mathcal{N}} { c_{kk,om}  \boldsymbol{w^{o}}^{T}E[\boldsymbol{w}_{m,i}]  } +\nu_{k} \boldsymbol{w^{o}}^{T}\boldsymbol{w^{o}}+M \mu_{k}^{2} \sigma^{2}_{u_k} \sigma^{2}_{v_k},
\end{align} where
\begin{align}
\label{coeff_general_N_nodes_2}
c_{kk,mm}&=\eta_{k} \sum_{j \in \mathcal{V}} {p_{j} a_{k,m,j}^2}, \\
\label{coeff_general_N_nodes_3}
c_{kk,mn}&=2 \eta_{k} \sum_{j \in \mathcal{V}} {p_{j} a_{k,m,j} a_{k,n,j},~ m \neq n}, \\
\label{coeff_general_N_nodes_kkom}
c_{kk,om}&=2 \sum_{j \in \mathcal{V}} {p_{j} a_{k,m,j}(\epsilon_{k}-\nu_{k})},  \\
\label{coeff_general_N_nodes_4}
c_{v_k} &= M \mu_{k}^{2} \sigma^{2}_{u_k} \sigma^{2}_{v_k},
\end{align}  and:
\begin{align}	
\label{coeff_general_kk}
\eta_{k} &=1-2 \mu_k \sigma^{2}_{u_k}+ \mu_k^2 (M+2) \sigma_{u_k}^4  ,\\
\label{coeff_general_kknu}
\nu_k &=\mu_k^2 (M+2) \sigma_{u_k}^4.
\end{align} Let $\mathcal{W}_{kl}$ denote the one-sided $z\text{-transform}$ of $E[\boldsymbol{w}^{T}_{k,i} \boldsymbol{w}_{l,i}]$. Taking the $z$-transform of both sides of (\ref{klgeneral_spatial}) and (\ref{kkgeneral_spatial}) and after some algebra we obtain 
\begin{align}
\label{kkgeneral_spatial_zdoamin_re}
(1-z^{-1} c_{kk,kk})\mathcal{W}_{kk}- z^{-1} \sum_{m \in \mathcal{N}, m \neq k} {c_{kk,mm} \mathcal{W}_{mm}} & 
- z^{-1} \sum_{m \in \mathcal{N}, m \neq k} {c_{kk,mn} \mathcal{W}_{mm}} \nonumber \\ &~~ =z^{-1} \sum_{m \in \mathcal{N}} {c_{kk,om} \boldsymbol{w^{o}}^{T} \mathcal{W}_{m}} + \frac{\nu_{k} \boldsymbol{w^{o}}^{T} \boldsymbol{w^{o}}+c_{v_{k}} }{1-z^{-1}}, 
\end{align}
\begin{align}
\label{klgeneral_spatial_zdoamin_re}
(1-z^{-1} c_{kl,kl})\mathcal{W}_{kl}-z^{-1} \sum_{m \in \mathcal{N} } {c_{kl,mm} \mathcal{W}_{mm}} & - z^{-1} \sum_{\substack{m,n \in \mathcal{N}, m \neq n \\ m \neq k, n \neq l}} {c_{kl,mn} \mathcal{W}_{mm}} \nonumber \\ &~~ = z^{-1} \sum_{m \in \mathcal{N}} {c_{kl,om} \boldsymbol{w^{o}}^{T} \mathcal{W}_{m}} + \frac{\nu_{kl} \boldsymbol{w^{o}}^{T} \boldsymbol{w^{o}} } {1-z^{-1}}. 
\end{align} Note that (\ref{kkgeneral_spatial_zdoamin_re}) and (\ref{klgeneral_spatial_zdoamin_re}) are the expressions at one-node level and two-node level that completely describe the coupling effects among different nodes in a diffusion estimation algorithm. The total number of equations in a network including $N$ nodes would then be $Q=\frac{N(N+1)}{2}$. In order to build the equations in a compact form, we consider writing all one-node level equations followed by those representing two-node level. Thus, ensuring that the permutation of the set of equations is selected as $\{11, \ldots, NN, 12,\ldots, 1N,23,\ldots,2N,\ldots, N-2~N-1,N-2~N, N-1~N\}$, the system description in $z$-domain can be represented as follows:

\begin{align}
\label{setofequ}
&
  \underbrace{\begin{bmatrix}
  1-z^{-1}c_{11,11} & -z^{-1}c_{11,22} & \cdots & -z^{-1}c_{11,N-1~N} \\
  -z^{-1}c_{22,11} & 1-z^{-1}c_{22,22} & \cdots & -z^{-1}c_{22,N-1~N} \\
  \vdots  & \vdots  & \ddots & \vdots  \\
  -z^{-1}c_{N-1~N,11} & -z^{-1}c_{N-1~N,22} & \cdots & 1-z^{-1}c_{N-1~N,N-1~N}
 \end{bmatrix}}_{\left[ \mathcal{C} \right] _{Q \times Q}}
 \underbrace{\begin{bmatrix}
 \mathcal{W}_{11}\\
 \mathcal{W}_{22}\\
 \vdots \\
 \mathcal{W}_{N-1~N}
 \end{bmatrix}}_{\left[ \mathcal{W} \right] _{Q \times 1}}
 = \\ \nonumber
&  \qquad \qquad \underbrace{\begin{bmatrix}
 z^{-1}\sum_{m \in \mathcal{N}} {c_{11,om} \boldsymbol{w^{o}}^{T} \mathcal{W}_{m}} + \frac{\nu_{1} \boldsymbol{w^{o}}^{T} \boldsymbol{w^{o}}+c_{v_{1}} } {1-z^{-1}} \\
  z^{-1}\sum_{m \in \mathcal{N}} {c_{22,om} \boldsymbol{w^{o}}^{T} \mathcal{W}_{m}} + \frac{\nu_{2} \boldsymbol{w^{o}}^{T} \boldsymbol{w^{o}}+c_{v_{2}} } {1-z^{-1}} \\
  \vdots \\
   z^{-1}\sum_{m \in \mathcal{N}} {c_{N-1~N,om} \boldsymbol{w^{o}}^{T} \mathcal{W}_{m}} + \frac{\nu_{n-1~n} \boldsymbol{w^{o}}^{T} \boldsymbol{w^{o}} }  {1-z^{-1}}
 \end{bmatrix}}_{\left[ \mathcal{D} \right] _{Q \times 1}}.
\end{align}

Denoting the $i^{th}$ column of $\mathcal{C}$ by $\mathcal{C}_{i},~ i=1,2,\ldots,Q$, we write
\begin{align}
\left[ \mathcal{C}_{1} ~\mathcal{C}_{2} ~\ldots ~\mathcal{C}_Q \right] _{Q \times Q} \left[ \mathcal{W} \right] &_{Q \times 1} =\left[ \mathcal{D} \right] _{Q \times 1}.
\end{align}

Recall that our objective is to find $\mathcal{W}_{kk},~k=1,2, \ldots, N$, i.e. the first $N$ elements of $\mathcal{W}$ from the set of equations described in (\ref{setofequ}). Let $\mathcal{C_{\mathcal{D}}}_{i},~i=1,2,\ldots,N$ denote the matrix obtained after replacing the $i^{th}$ column of $\mathcal{C}$ by $\mathcal{D}$. Then, using Cramer's rule we obtain  $\mathcal{W}_{kk}$ as follows:

\begin{equation}
\label{wkk}
\mathcal{W}_{kk}=\frac{det(\mathcal{C_{\mathcal{D}}}_{k})}{det(\mathcal{C})},~k=1,2, \ldots, N.
\end{equation}

We can rewrite matrix $\mathcal{D}$ as follows:
\begin{align}
\label{ddwo}
\mathcal{D}=\underbrace{\begin{bmatrix}
 z^{-1}\sum_{m \in \mathcal{N}} {c_{11,om} \boldsymbol{w^{o}}^{T} \mathcal{W}_{m}} + \frac{\nu_{1} \boldsymbol{w^{o}}^{T} \boldsymbol{w^{o}} } {1-z^{-1}} \\
  z^{-1}\sum_{m \in \mathcal{N}} {c_{22,om} \boldsymbol{w^{o}}^{T} \mathcal{W}_{m}} + \frac{\nu_{2} \boldsymbol{w^{o}}^{T} \boldsymbol{w^{o}} } {1-z^{-1}} \\
  \vdots \\
   z^{-1}\sum_{m \in \mathcal{N}} {c_{N-1~N,om} \boldsymbol{w^{o}}^{T} \mathcal{W}_{m}} + \frac{\nu_{N-1~N} \boldsymbol{w^{o}}^{T} \boldsymbol{w^{o}} }  {1-z^{-1}}
 \end{bmatrix}}_{\left[ \mathcal{D}_{w^o} \right]} +
 \underbrace{\begin{bmatrix}
 \frac{c_{v_{1}} } {1-z^{-1}} \\
  \frac{c_{v_{2}} } {1-z^{-1}} \\
  \vdots \\
  0 
 \end{bmatrix}}_{\left[ \mathcal{D}_{v} \right] }.
\end{align} It follows that
\begin{align}
\label{dsplit}
det(\mathcal{C_{\mathcal{D}}}_{i})=\begin{vmatrix} \mathcal{C}_1~\cdots~\mathcal{D}_{w^o}~\mathcal{C}_Q\end{vmatrix}+\begin{vmatrix}
\mathcal{C}_1~\cdots~\mathcal{D}_{v}~\mathcal{C}_Q\end{vmatrix}.
\end{align}
Using (\ref{dsplit}) in (\ref{wkk}), we obtain the following expression
\begin{align}
\label{wkksplit}
\mathcal{W}_{kk}=\frac{\begin{vmatrix}
\mathcal{C}_1~\cdots~\mathcal{D}_{w^o}~\mathcal{C}_Q\end{vmatrix}}{\begin{vmatrix}\mathcal{C}\end{vmatrix}}+\frac{\begin{vmatrix}
\mathcal{C}_1~\cdots~\mathcal{D}_{v}~\mathcal{C}_Q\end{vmatrix}}{\begin{vmatrix}\mathcal{C}\end{vmatrix}},
\end{align}
for all $k \in \mathcal{N}$. To proceed, we define the square matrix $\mathcal{C}^\prime$ of size $Q$ as $\mathcal{C}^\prime=[c_{kl,mn}]_{Q \times Q}$ with $c_{kl,mn}$ obtained from (\ref{coeff_general_N_nodes}),(\ref{coeff_general_N_nodes_1}),(\ref{coeff_general_N_nodes_2}) and (\ref{coeff_general_N_nodes_3}) and arranged as in $\mathcal{C}$ in (\ref{setofequ}). To simplify, we define $c_s$, $c_{{w^o},s}$ and $c_{{v_k},s}, k=1,2,\ldots,N$ at the steady-sate as follows:

\begin{eqnarray}
\label{cvs}
c_{{v_k},s}&=&\lim_{z \to 1} {(z-1)\begin{vmatrix}
\mathcal{C}_1~\mathcal{C}_2~\cdots~\mathcal{D}_{v}~\mathcal{C}_Q\end{vmatrix}} \nonumber \\
&=&\begin{vmatrix}
  1-c_{11,11} & -c_{11,22} & \cdots &c_{v_{1}} & -c_{11,N-1~N} \\
  -c_{22,11} & 1-c_{22,22} & \cdots &c_{v_{2}} & -c_{22,N-1~N} \\
  \vdots  & \vdots  & \ddots & &\vdots \vdots  \\
  -c_{N-1~N,11} & -c_{N-1~N,22} & \cdots &0 & 1-c_{N-1~N,N-1~N}
 \end{vmatrix}.
\end{eqnarray}

\begin{align}
\label{cs}
&c_{s}=\lim_{z \to 1} {\begin{vmatrix}
\mathcal{C}\end{vmatrix}} = \begin{vmatrix} \mathcal{I}_N - \mathcal{C}^\prime \end{vmatrix},
\end{align} \begin{align}
\label{cwos}
c_{{w^o},s}&=\lim_{z \to 1} {(z-1)\begin{vmatrix}
\mathcal{C}_1~\cdots~\mathcal{D}_{w^o}~\mathcal{C}_Q\end{vmatrix}}=\boldsymbol{w^{o}}^{T}\boldsymbol{w^{o}} \begin{vmatrix}\mathcal{C}\end{vmatrix}.
\end{align}
Notice that the derivation of (\ref{cwos}) is given in Appendix~\ref{app_ms}. Let $w_{k,s}$ denote the expectation of the norm of the weight vector corresponding to the $k^{th}$ node at the steady-state as expressed in (\ref{msdvalues}). Then, using the final value theorem, $\mathcal{W}_{kk}$ as described in (\ref{wkksplit}) and the steady-state quantities as defined in (\ref{cvs}) and (\ref{cs}), we arrive at the following result: 
\begin{align}
\label{wks_app_ms}
w_{k,s} & \triangleq  \lim_{i \to \infty } {E[ \boldsymbol{w}_{k}^{T} \boldsymbol{w}_{k} ]} \nonumber \\ 
&=\lim_{ z \to 1 } { (z-1) \mathcal{W}_{kk} }  = \boldsymbol{w^{o}}^{T}\boldsymbol{w^{o}}+ \frac{c_{{v_k},s}}{c_s}.
\end{align}

Finally, the closed-form expression for local $\text{MSD}_{k}$ is given by
\begin{equation}
\label{colsedformmsd}
\text{MSD}_k=\frac{c_{{v_k},s}}{c_s}.
\end{equation}

We notice that the global network MSD is obtained by averaging over the local MSD's as follows:
\begin{equation}
\label{msdglobal}
\text{MSD}=\frac{1}{N} \sum_{k \in \mathcal{N}} {\text{MSD}_{k}}.
\end{equation}

\subsection{Mean-square Stability}
\label{Mean-square Stability}
In this subsection, we discuss the mean-square stability of diffusion LMS algorithms in the presence of transmission errors. In particular, our aim is to answer the following questions. How do transmission errors affect the convergence of diffusion algorithms in the mean-square sense? Is there an explicit sufficient condition to ensure mean-square stability? To address these important issues, we use the same approach as the one presented in Subsection \ref{Mean Stability}. 

We start with arguing that in order to use Cramer's law in (\ref{setofequ}) around $\abs{z} \to 1$, we must provide condition to prevent any root of $det(\mathcal{C})=0$ be placed on the unit circle. Furthermore, all of the corresponding roots must lie within the unit circle to guarantee stability in the mean-square sense. We note that these roots are the eigenvalues of $\mathcal{C}^\prime$. For convenience, we rewrite matrix $\mathcal{C}^\prime$ as $\mathcal{C}^\prime=[\gamma_{i,j}]_{Q \times Q}$ and define $\eta^\prime_i, i=1,2,\ldots,Q$ to denote the common factor ($\eta_k$ in (\ref{coeff_general_kk}) and $\eta_{kl}$ in  and (\ref{coeff_general_kl})) of elements in the $i^{th}$ row of $\mathcal{C}^\prime$. Let us define $\lambda_k^\prime,~ k=1,2,\ldots,Q$ to refer to the eigenvalues of  $\mathcal{C}^\prime$. Using (\ref{sumis1}), (\ref{coeff_general_N_nodes}),(\ref{coeff_general_N_nodes_1}),(\ref{coeff_general_N_nodes_2}) and (\ref{coeff_general_N_nodes_3}) and noting the structure of $\mathcal{C}^\prime$, we find one important feature of the rows of $\mathcal{C}^\prime$
\begin{equation}
\label{meansquaresumofm}
\sum_{j=1}^Q {\gamma_{i,j}}=\eta^\prime_i, \quad i=1,2,\ldots,Q.
\end{equation}

Using this result and considering the induced infinity-norm of matrix  $\mathcal{C}^\prime$ and the similar principles already discussed in Subsection \ref{Mean Stability}, we find out that to satisfy $\max_{1 \leq k \leq Q} \abs{\lambda_k^\prime} < 1$, it is sufficient to ensure that $\abs{\eta^\prime_i} < 1,~i=1,2,\ldots,Q$, or equivalently 
\begin{equation}
\label{meansquarecondition}
\abs{\eta_{k}} < 1,~k\in \mathcal{N},\quad \text{and} \quad \abs{\eta_{kl}} < 1,~k,l\in\mathcal{N}, k \ne l.
\end{equation} 

It is worth mentioning that (\ref{coeff_general_kk}) suggests that $\eta_k$ only depends on the local step size of the individual filter of a single node and the statistics of the filter input. On the other hand, from (\ref{coeff_general_kl}), it is evident that $\eta_{kl}$ accounts for the interaction among node pairs and is a function of step sizes and the statistics of the inputs at two different nodes. The solution of (\ref{meansquarecondition}) provides useful and practically applicable lower and upper bounds for step sizes which can be written as follows:
  
\begin{equation}
\label{meansquareconditionmu}
0 < \mu_k < \frac{2}{(M+2)\sigma^2_{u_k}}, \quad k=1,2,\ldots ,N.
\end{equation}

The following important result is drawn from the above discussions. In a network with $N$ nodes using distributed diffusion estimation algorithm (\ref{nodek})-(\ref{nodekd}) with combining weight matrices satisfying (\ref{constraintona}), in the presence of multiplicative transmission errors modeled as (\ref{modeltranserror}), the sufficient condition provided by (\ref{meansquareconditionmu}) ensures stability in the mean-square sense. An important feature of the condition (\ref{meansquareconditionmu}) is that it is not dependent on error probabilities $p_{k,l}$. This suggests that transmission errors do not lead to the mean-square divergence of diffusion estimation algorithms.

It is worth mentioning that condition (\ref{meansquareconditionmu}) represents a novel bound even in the case of perfect information exchange. Considering perfect transmissions, it has been argued that sufficiently small step-sizes that satisfy mean stability, will also ensure mean-square stability \cite{Cattivelli2010}. Sufficiently small step-sizes may however be a conservative approach which leads to slow convergence rates and thus high energy requirements at individual nodes. This is undesirable given that in WSNs energy is crucially scarce. Furthermore, in certain applications where minimizing the speed of convergence is more important than achieving a small steady-state error one should not select very small step-sizes during the transient time. All these reveal the importance of the upper bound of step-sizes provided by (\ref{meansquareconditionmu}). Another practical significance of (\ref{meansquareconditionmu}) is that the condition is fully distributed, i.e, each node can locally select its step-size according to the statistics of the filter input.  

\section{Modeling transmission errors}
\label{Modeling transmission errors}

As previously discussed, there exist various uncertainties in WSNs such as  mobility of nodes and time-varying network topology, interference and multipath fading, signal attenuation at the PHY layer, packet loss at the medium access control (MAC), etc. Any of these uncertainties may result in transmission errors during information exchange. Thus, in order to accurately evaluate transmission errors in WSNs, it is required to consider all sources of errors depending on the type of application. Given the scope of this paper, we limit our discussion to the MAC-level errors as one example of how to quantify $p_{k,l}$ throughout the rest of this section. 

There is a vast literature on designing efficient channel access mechanisms for WSNs and due to page restrictions, we refer the reader to \cite{demirkol2006,injong2008,bachir2010} for further details. For our purpose it is sufficient to concentrate on the exponential backoff procedure that has been standardized as the basic access mechanism in IEEE 802.11 \cite{802.11} and the contention phase of the IEEE 802.15.4 \cite{802.15.4} which is designed for low rate WSNs. In this mechanism, every node that has a packet to transmit senses the channel and if it is idle for a period called distributed interframe space (DIFS), the node transmits. Otherwise, it waits until the channel is idle for a DIFS and then starts a backoff. The random backoff period is uniformly selected between 0 and the contention window. The initial size of the contention window is $CW$ and is doubled at each retransmission. Let $R$ denotes the maximum number of retransmissions. Then, the maximum contention window size is $CW_{\text{max}}=2^R CW$. The backoff counter is decremented after each slot time provided that the channel is sensed idle. The transmission starts when the backoff counter is zero. If an acknowledgement (ACK) is received from the destination, the transmission is successful; otherwise, a collision is inferred. Let $q_{k,l}~k \in \mathcal{N},~l \in \mathcal{N}_k$ denote the probability of collision assigned to the transmissions with node $k$ as destination and node $l$ as source. Furthermore, we assume that each node $k$ has $n_k$ neighbors (degree of node $k$ is $n_k$). Node $k$ successfully receives a packet from node $l$ if none of its remaining neighbors or itself transmit simultaneously. We assume that all nodes are deploying the same set up for backoff procedure, i.e., the maximum number of retransmissions and the initial window sizes are identical. Consequently, each node transmits a packet with probability $\tau$. Thus, the collision probability $q_{k,l}$ can be written as follows:
\begin{equation}
\label{collision_link}
q_{k,l}=1-(1-\tau)^{n_k} \quad k \in \mathcal{N},~l \in \mathcal{N}_k.
\end{equation}

We assume that each node knows how many neighboring nodes it has; thus, $n_k$ is known. Additionally, we consider the seminal paper of Bianchi  that develops a two state Markov chain to evaluate the performance of the exponential backoff algorithm to express $\tau$ as a function of $R$, $CW$ and $p_{k,l}$ as follows \cite{bianchi2000}:
\begin{equation}
\label{transmission_tau}
\tau=\frac{2(1-2 q_{k,l})} {(1-2 q_{k,l})(CW+1)+q_{k,l} CW (1-(2 q_{k,l})^R) }
\end{equation}      

Solving (\ref{collision_link}) and (\ref{transmission_tau}), we find the probability of collision on each link. In a channel access mechanism based on the exponential backoff procedure as described above, packet collisions are closely related to the transmission errors $p_{k,l}$. If a packet collides more than the maximum number of retransmissions during the information exchange period, then the packet is discarded and a transmission error occurs. In other words, transmission errors $p_{k,l}$ on all directional links to node $k$ are identical and equivalent to the packet loss experienced by node $k$ which is denoted by $q_k$ for convenience. More precisely, we have
\begin{equation}
\label{loss probability}
p_{k,l}=q_{k,l}^{m+1}=q_k, \quad k \in \mathcal{N}, l \in \mathcal{N}_k.
\end{equation}  

\section{Performance improvement}
\label{Performance improvement}
Throughout this section, we briefly review different combining rules and then discuss how to use the knowledge of the properties of transmission errors in order to improve the performance of diffusion algorithms. Different methods for combining the received information from neighboring nodes can be divided into three groups. The first group includes the methods that solely rely on the information regarding each node's degree such as relative degree \cite{Cattivelli2008}, uniform \cite{blondel2005}, Laplacian \cite{scherber2004}, \cite{xiao04} and Metropolis \cite{xiao04}, as well as the maximum degree method \cite{xiao05}  which only uses the information regarding the total number of nodes in the network. We notice that in these methods no information regarding the noise ($v_{k,i}$ in (\ref{nodekd})) variances across the network is required. The second group includes the rules that only takes into account the noise levels across the network such as the relative variance method. In the relative variance method, larger weight is assigned to the node with smaller noise variance. Finally, the third group includes the methods that require information regarding both the nodes's degree and noise levels such as relative degree variance rule \cite{Cattivelli2010} and Hasting rule \cite{zhao12}. If any of the methods in the last two groups are used, and unlike the first group, there should be mechanisms to estimate the local noise variance and then distribute such information across the network . This extra cost will however result in lower MSD compared to the methods of the first group. The combining rules constructed only in terms of the degrees of the nodes are considerably more efficient in networks where the noise levels experienced by the neighboring nodes are in the same range. In such networks, it is rational to assign more weight to the nodes with larger degrees. In particular, the relative degree policy constructs the mixing coefficients in the following manner:
\begin{equation}
\label{relative degree}
a_{k,l,i}=\frac{n_{l}}{\sum_{m \in \mathcal{S}_{k,i}}{n_{m}}}, ~ l \in \mathcal{S}_{k,i}, ~ \text{(relative degree rule)}, \nonumber
\end{equation} where $n_l$ is the cardinality of $\mathcal{N}_l$. In the presence of transmission errors, it is reasonable to modify the relative degree rule to account for errors. It is evident that such errors affect the connectivity of the network and in particular the degrees of nodes at each iteration of the adaptive estimation algorithm. The most natural idea is to take into account the transmission errors to define the true degree of each node. Particularly, we obtain the effective degree of each node denoted by $n^\prime_k$ with multiplying its degree when no error occurs by its average probability of successful reception:
\begin{equation}
\label{true degree}
n^\prime_k=n_k (1-q_k), \quad k \in \mathcal{N}.
\end{equation}
In general, the average error probability $q_k$ can be calculated at each node by counting the successfully received information packets over a period. The average error probability when concentrating on the MAC-level errors as described in Section \ref{Modeling transmission errors}, is the average packet loss probability experienced by each node which can be locally estimated over a short period and based on the number of the received ACKs. It can also for instance be computed according to the model we already considered for the backoff procedure in Section \ref{Modeling transmission errors}. The new combining rule which we refer to as the \textit{enhanced relative degree method} can be expressed as follows:
\begin{align}
\label{enhanced relative degree}
a_{k,l,i}=\frac{ n^\prime_{l}} {\sum\limits_{m \in \mathcal{S}_{k,i}}{n^\prime_{m}}}, ~ l \in \mathcal{S}_{k,i}, ~ \text{(enhanced relative degree rule)}. \nonumber
\end{align}

\section{Simulations}
\label{sim}
In the first phase of simulations, we consider identical failure probabilities. To justify the assumption of identical error probabilities, consider as an example a scenario where each node randomly turns into sleep mode for power saving and does not share information with its neighbors. For such source of failure, it is reasonable to assume identical error probabilities in a network with homogeneous nodes. In the second phase of simulations, we consider another practical scenario where errors occur at the MAC level while nodes access the channel using a backoff procedure. In such scenario, we use (\ref{collision_link})-(\ref{loss probability}) to model non-identical error probabilities experienced by each node.
\subsection{Uniform Error Probabilities}
\begin{figure}[htbp]
\centering
\includegraphics [width=0.5\columnwidth] {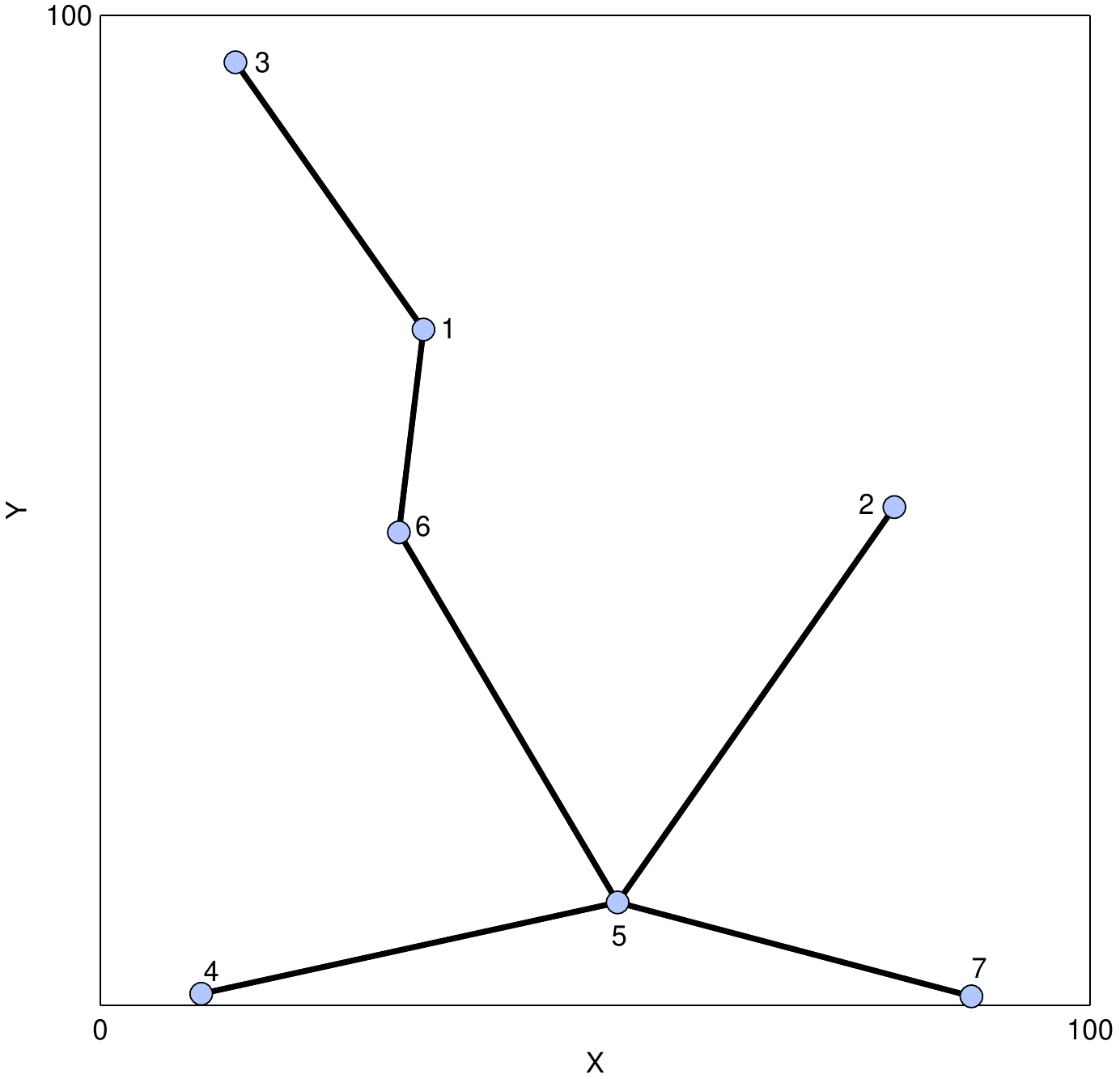}
\caption {Network topology for 7-node network. The node index is shown next to each node.}
\label{topo_7.pdf}
\end{figure}

We consider a small 7-node network where nodes are randomly distributed in a square area with side $S=100$ units. There exists a link between any pair of nodes with a distance less than $50$ units. The network topology is shown in Fig.~\ref{topo_7.pdf}. For convenience, we denote the error probability by $p$ since through this simulation, transmission errors are assumed to be identical for all links, i.e., $p_{k,l}=p, (k,l)\in \mathcal{L}$. 
\begin{figure}[htbp]
\centering
\includegraphics [width=0.5\columnwidth] {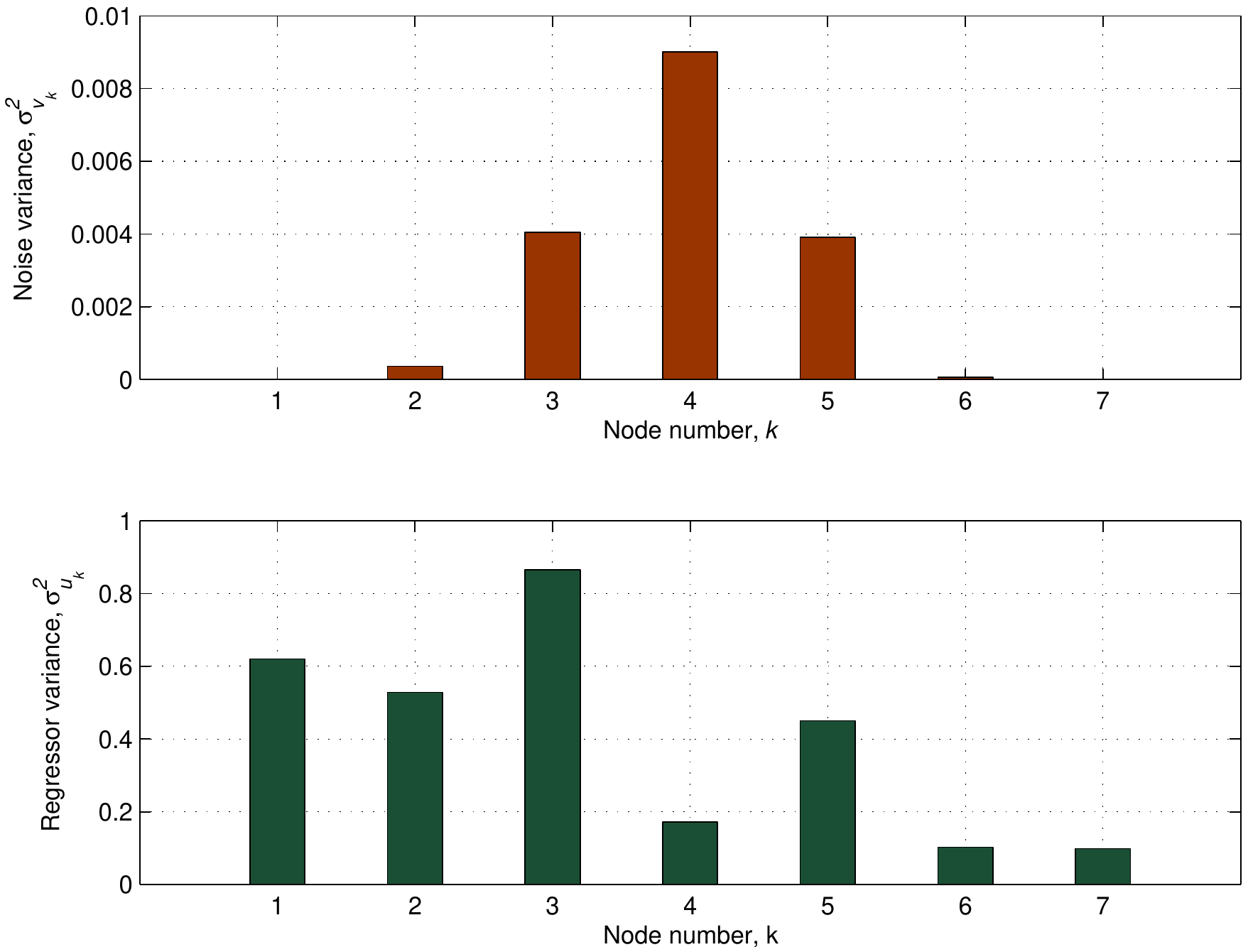}
\caption {Noise power profile $\sigma^2_{v_k}$ (top) and regressor power profile $\sigma^2_{u_k}$ (bottom) for 7-node network in Fig.~\ref{topo_7.pdf}.}
\label{7_profile.pdf}
\end{figure}
\begin{figure}[htbp]
\centering
\includegraphics[width=0.5\columnwidth] {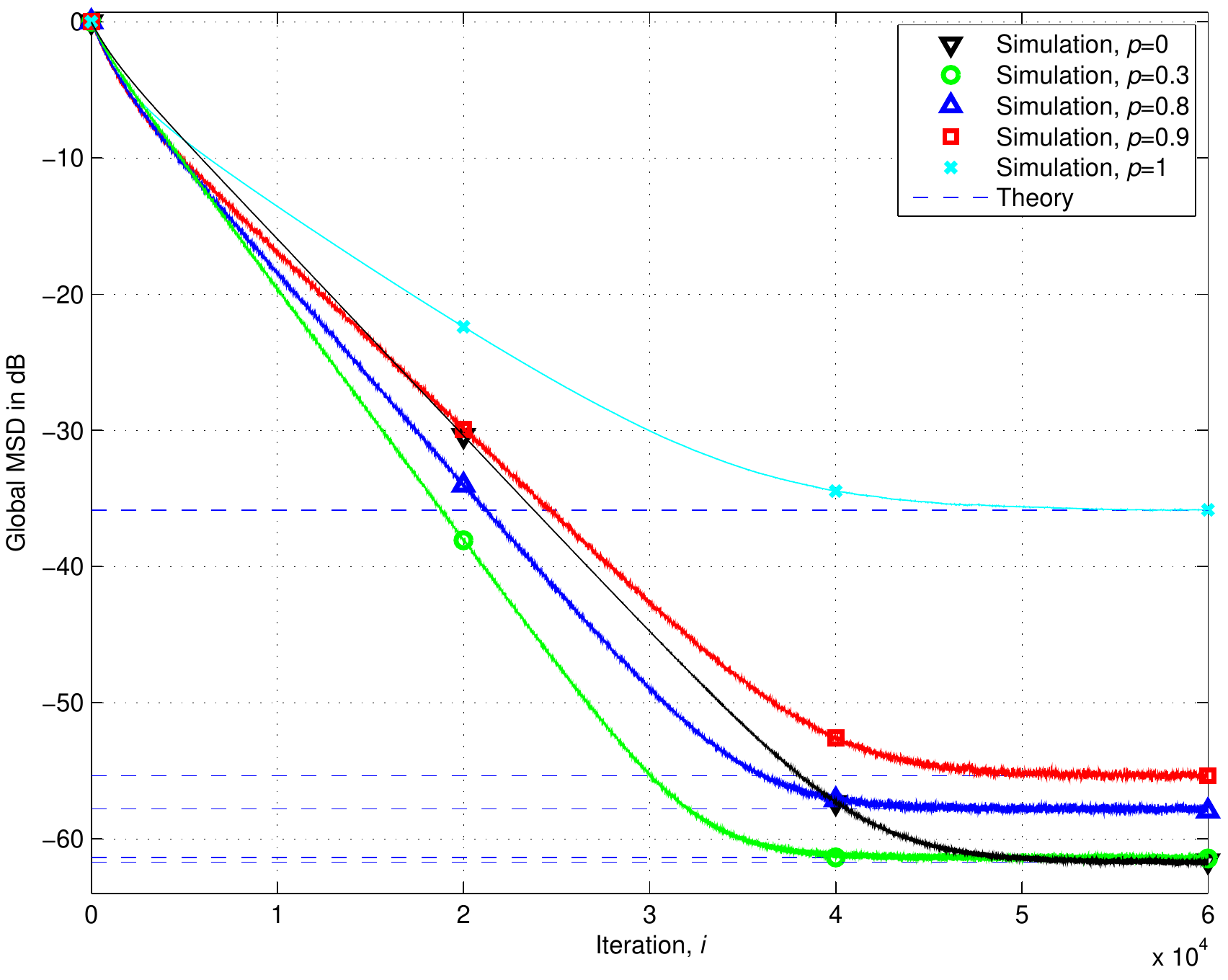}
\caption {Learning curve in terms of the global MSD in dB for different values of error probability $p \in \{0,0.3,0.8,0.9,1\}$ in 7-node network. The dashed lines show the theoretical expression (\ref{msdglobal}) for the steady-state MSD.}
\label{learn_7_p_id_org_paper.pdf}
\end{figure}
Without loss of generality, we apply the relative variance rule which gives more weight to nodes with lower noise variance to obtain the mixing coefficients \cite{Cattivelli2010}. Hence, it follows that $a_{k,l,i}=\frac{\sigma_{v_l}^{-2}}{\sum_{m \in \mathcal{S}_{k,i}}{\sigma_{v_m}^{-2}}}$ for $l \in \mathcal{S}_{k,i}$ and otherwise we have that $a_{k,l,i}=0$. Notice that any rule for finding the mixing coefficients is possible as long as it satisfies the condition $\sum_{l\in\mathcal{N}}{a_{k,l,j}}=1,~k=1,2,\ldots,N$ for all $j$, as discussed in Section~\ref{sec:prob_form}. Each node has access to the distorted and noisy version of the same unknown vector $\boldsymbol{w}^{o}=\text{col}\{1,1,\ldots,1\}/\sqrt{M}$, with $M=200$ according to (\ref{nodekd}). The $M$-dimensional input regressors are assumed to be temporally independent Gaussian, but spatially correlated. The spatial correlation index $\rho_{kl}$ between two nodes $k$ and $l$ is obtained according to $\rho_{kl}=\rho^{|k-l|}$, where $\rho$ is a constant that lies between 0 and 1. The measurement noise is assumed to be white and Gaussian. The noise variances are generated randomly from $[10^{-6},10^{-2}]$ and shown in Fig.~\ref{7_profile.pdf} (top). The variances of input regressors are randomly selected over $(0,1]$ and depicted in Fig.~\ref{7_profile.pdf} (bottom). For all nodes we choose identical step-sizes, i.e., $\mu_{k}=\mu=0.001$. In order to obtain the performance measures, the results are averaged over 150 independent experiments each using $100,000$ iterations. A random noise is generated at each run according to the noise profile shown in Fig.~\ref{7_profile.pdf} (top). Fig.~\ref{learn_7_p_id_org_paper.pdf} shows the learning curves in terms of the global MSD for different values of error probability, i.e., $p \in \{0,0.3,0.8,0.9,1\}$. We observe that when the transmissions experience a high error, i.e., $p=0.9$, the global network MSD at the steady-state increases. Another observation is that the convergence speed of the diffusion estimation algorithm decreases as the transmission unreliability considerably increases, i.e., $p=0.9$. However, the global convergence rate might increase when the error probability increases. In other words, transmission errors might prevent the negative effect of a slow node on a fast converging node and thus improve the global convergence rate. In this network, for example, it might be better for node 7 with a low noise level to discard information received from node 5 with a high noise level rather than giving it some weight according to the relative variance rule. The policy of discarding such worthless received information is equivalent to not receiving the information at all due to transmission errors. 
\begin{figure}[htbp]
\centering
\includegraphics[width=0.5\columnwidth] {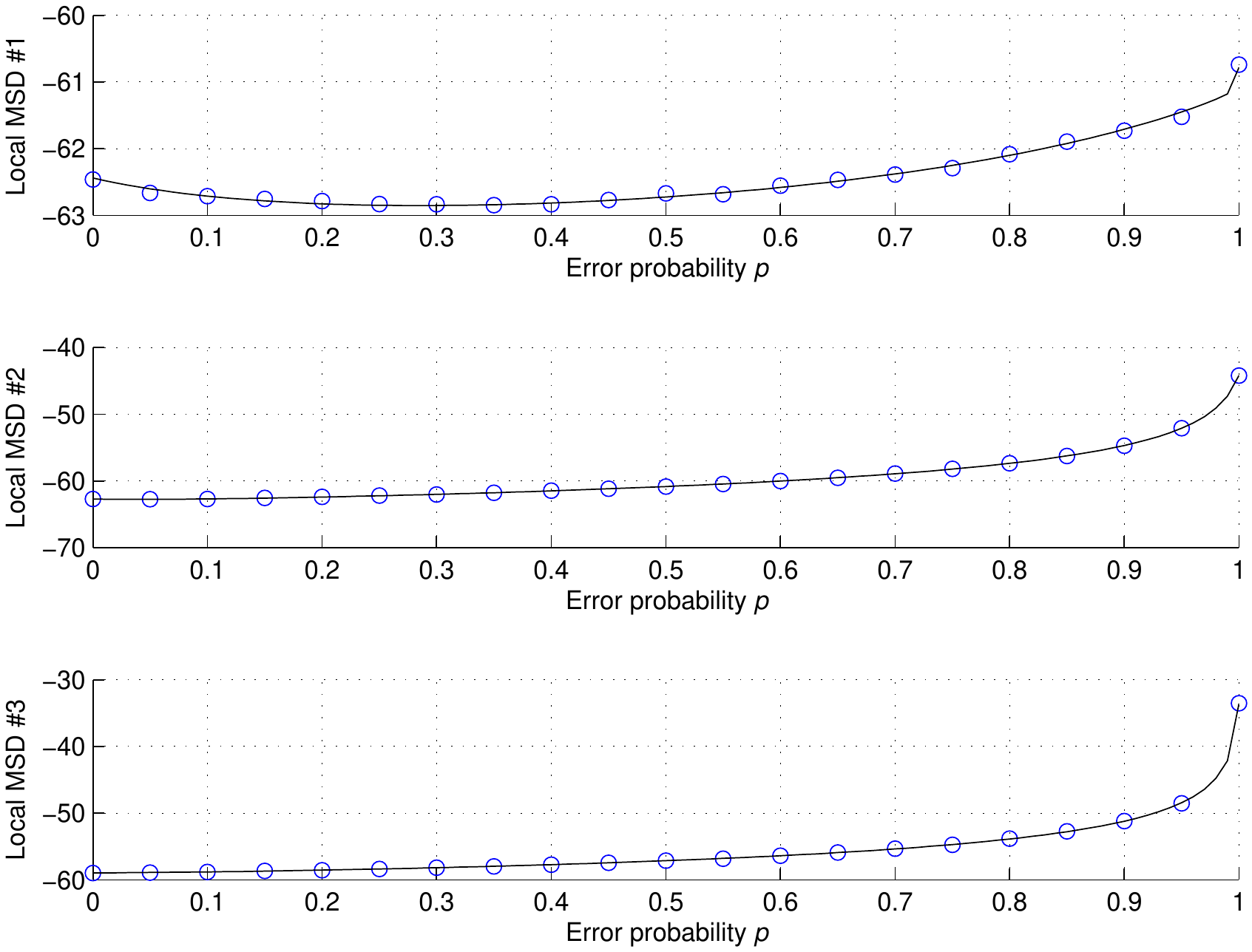}
\caption {Local steady-state MSD (dB) for nodes $\{1,2,3\}$ as a function of error probability in 7-node network. The solid line curves show the theoretical expression (\ref{colsedformmsd}) and the markers represent the simulation results.}
\label{2theo_N_7_link_6_rho_symm_local_13.pdf}
\end{figure}
\begin{figure}[htbp]
\centering
\includegraphics [width=0.5\columnwidth] {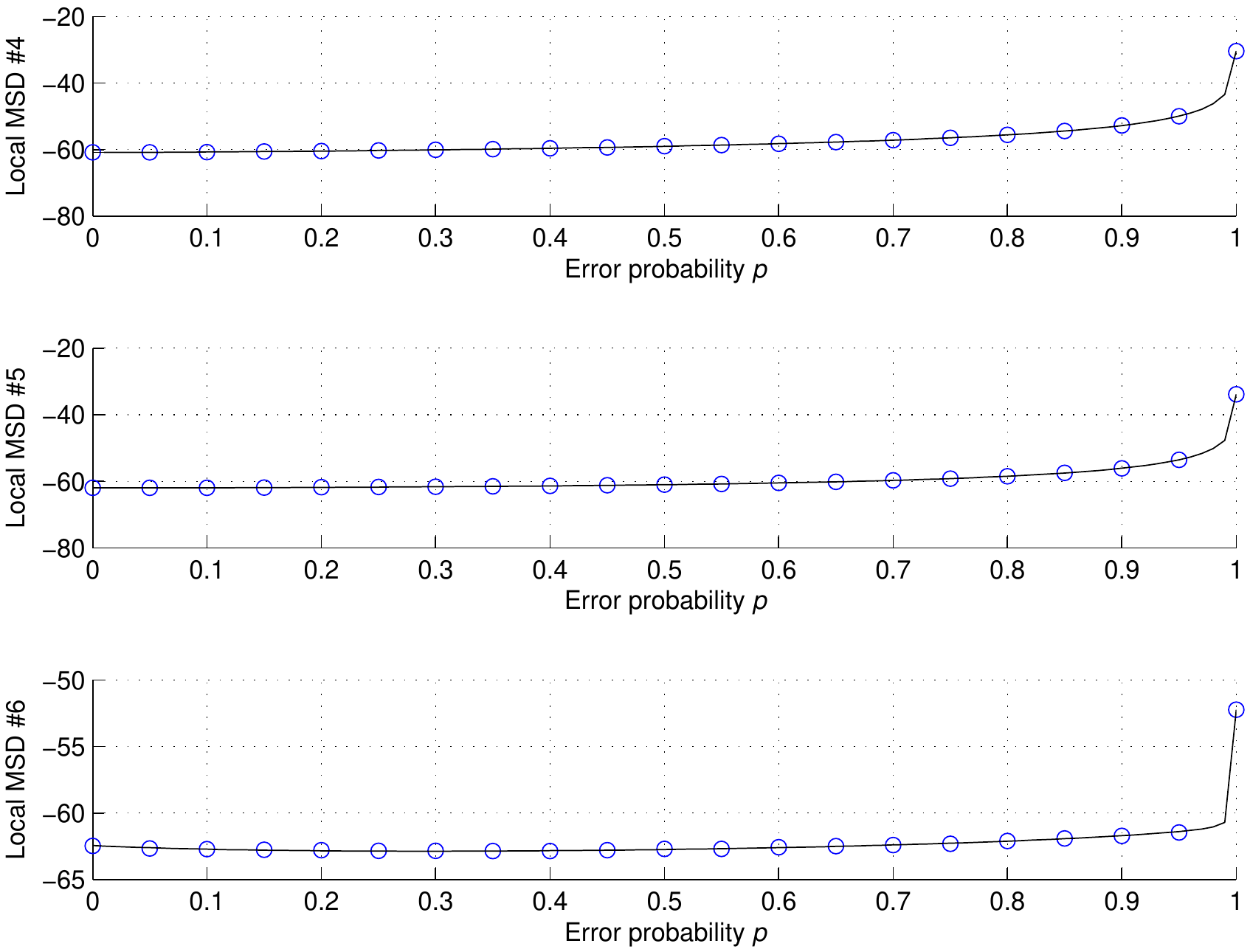}
\caption {Local steady-state MSD (dB) for nodes $\{4,5,6\}$ as a function of error probability in 7-node network. The solid line curves show the theoretical expression (\ref{colsedformmsd}) and the markers represent the simulation results.}
\label{2theo_N_7_link_6_rho_symm_local_46.pdf}
\end{figure}
\begin{figure}[htbp]
\centering
\includegraphics [width=0.5\columnwidth] {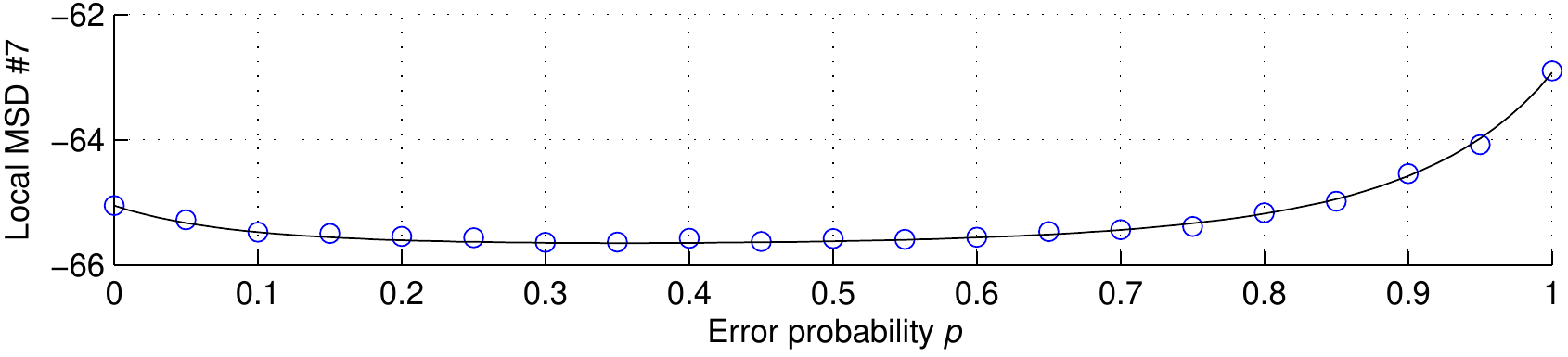}
\caption {Local steady-state MSD (dB) for node 7 as a function of error probability in 7-node network. The solid line curves show the theoretical expression (\ref{colsedformmsd}) and the markers represent the simulation results.}
\label{2_rep_new_local_7.pdf}
\end{figure}
To further investigate the impacts of the errors on the performance of the diffusion estimation algorithm, we evaluate the local steady-state MSD associated to each node by averaging over the last 1,000 samples of the individual learning curves. Local MSD curves as a function of error probability are illustrated in Figs.~\ref{2theo_N_7_link_6_rho_symm_local_13.pdf}-\ref{2_rep_new_local_7.pdf}. We observe that MSD curves obtained from analysis and simulation coincide with each other. It is seen that the MSD curves for a group of nodes $\{2,3,4,5\}$ are monotonically increasing functions of the error probability; on the contrary, for other nodes $\{1,6,7\}$ there exist non-zero values of probability of error that minimizes the corresponding local MSD. This is due to the fact that the noise variances of nodes $\{2,3,4,5\}$ are significantly higher than those of $\{1,6,7\}$. As a result, there exist certain optimum points in terms of the error probability that minimizes the noise amplification effect and consequently minimizes the local steady-state MSD of the nodes with low noise variances. Thus, we conclude that in general the local and global steady-state MSD curves are not necessarily  monotonically increasing functions of the error probability. 
\begin{figure}[htbp]
\centering
\includegraphics [width=0.5\columnwidth] {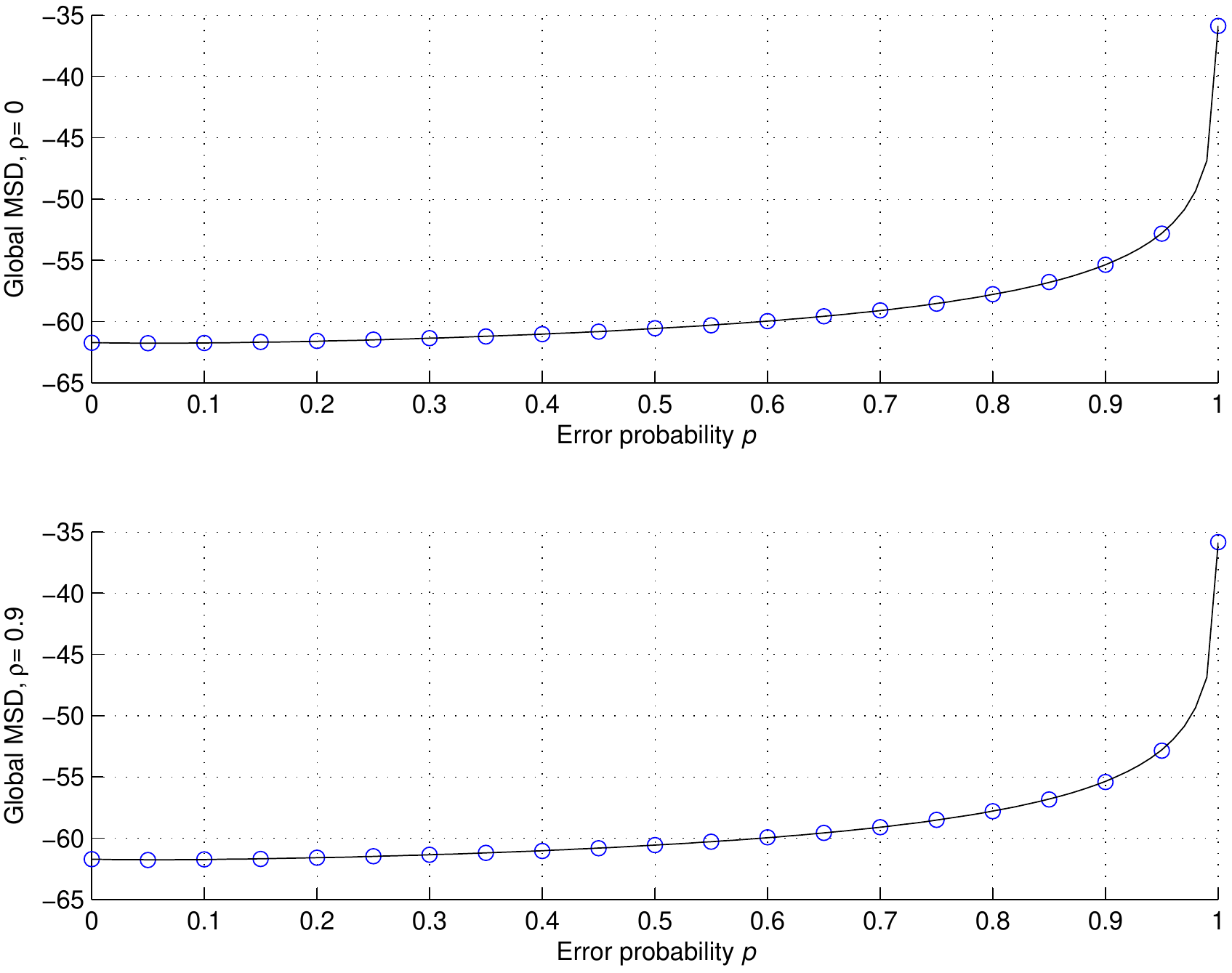}
\caption {Global steady-state MSD in dB as a function of error probability in 7-node network for different values of spatial correlation index when $\rho \in \{0,0.9\}$. The solid line curves show the theoretical expression (\ref{msdglobal}) and the markers represent the simulation results.}
\label{2_rep_avg.pdf}
\end{figure}
Fig.~\ref{2_rep_avg.pdf} shows the network global MSD curve for two different values of $\rho$. Note that in both scenarios, the minimum value of the steaty-state network MSD is not obtained at $p=0$ which confirms the non-increasing behavior of the global MSD curve. We observe that for both scenarios the minimum MSD is obtained at $p=0.06$ and the impact of the spatially correlated observation on the global steady-state MSD is negligible. In Table~\ref{1_node number according to the increasing order of local MSD}, we sort the nodes in an ascending order in terms of the local MSD. We observe that while all nodes achieve a lower MSD compared to non-cooperative mode by employing diffusion algorithm, the order of nodes changes depending on which region the network operates in, in terms of the transmission errors. In particular, for $p=0.9$, node 5 achieves a better performance compared to node 2. We can get insight into this behavior by noticing that the degree of node 5 is 4 while the degree of node 2 is 1. Consequently, even a small amount of information flow in the case of high error probability namely, $p=0.9$ significantly improves the performance of node 5. Meanwhile, as the error probability decreases, i.e., $p \rightarrow 0$, the degree of a node will not be a dominant factor for local performance improvement and node 2 achieves a lower MSD compared to node 5 due to its lower noise variance. 
\begin{table}[htbp]
\renewcommand{\arraystretch}{0.9}
\caption{node number according to the increasing order of local MSD  }
\label{1_node number according to the increasing order of local MSD}
	\centering
		\begin{tabular}{|c|c|}
		\hline
Error probability $p$ & node number\\
\hline
1 & 7 1 6 2 5 3 4 \\
\hline
0.9 & 7 1 6 5 2 4 3 \\
\hline
0.5 & 7 6 1 5 2 4 3\\
\hline
0.1 & 7 6 1 2 5 4 3\\ 
\hline
0 & 7 2 6 1 5 4 3\\
\hline
\end{tabular}
\end{table}  

\subsection{MAC-collision Based Errors}
\begin{figure}[htbp]
\centering
\includegraphics [width=0.5\columnwidth] {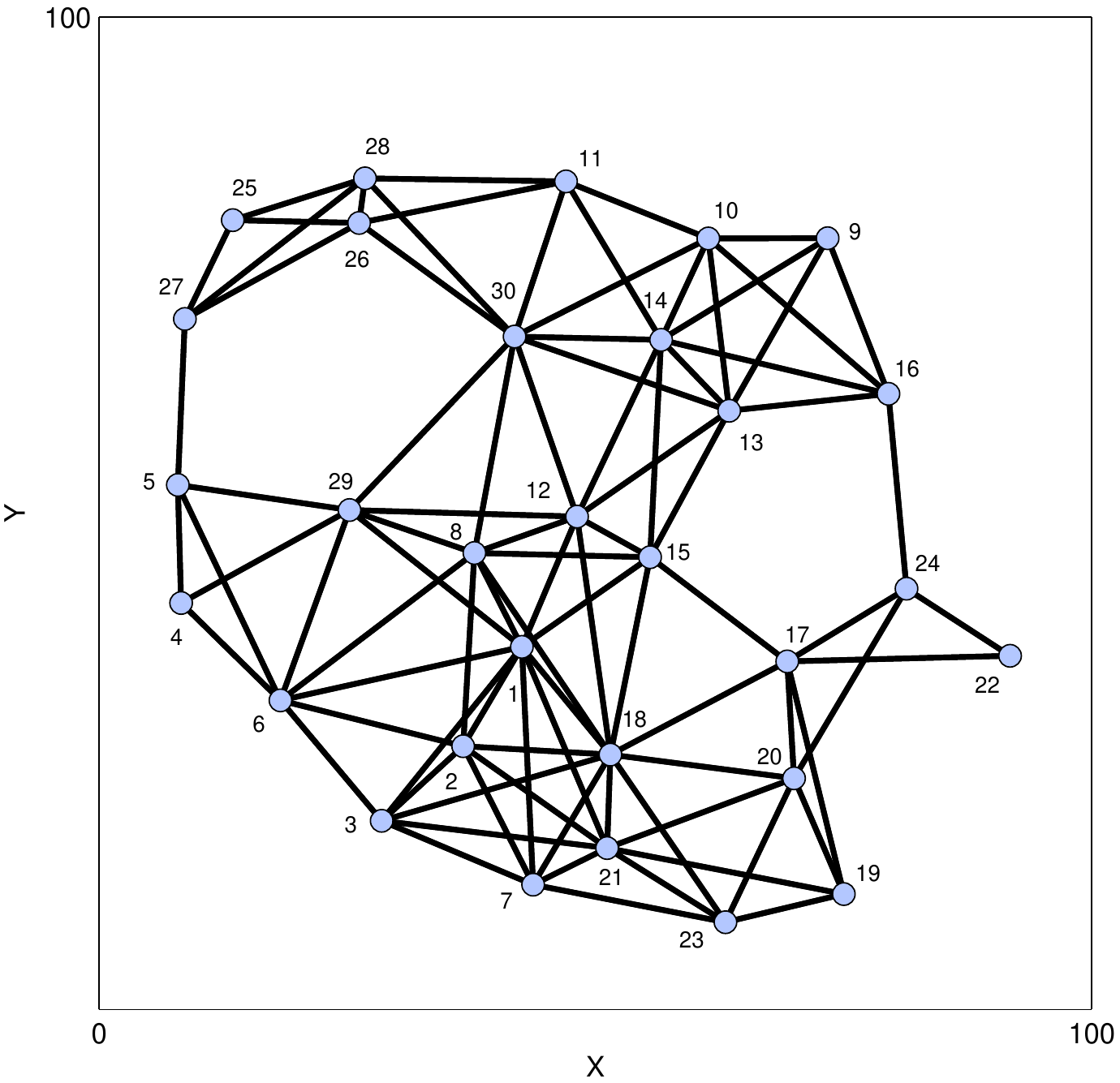}
\caption {Network topology for 30-node network. The node index is shown next to each node.}
\label{topo_30_dense.pdf}
\end{figure}
\begin{figure}[htbp]
\centering
\includegraphics [width=0.5\columnwidth] {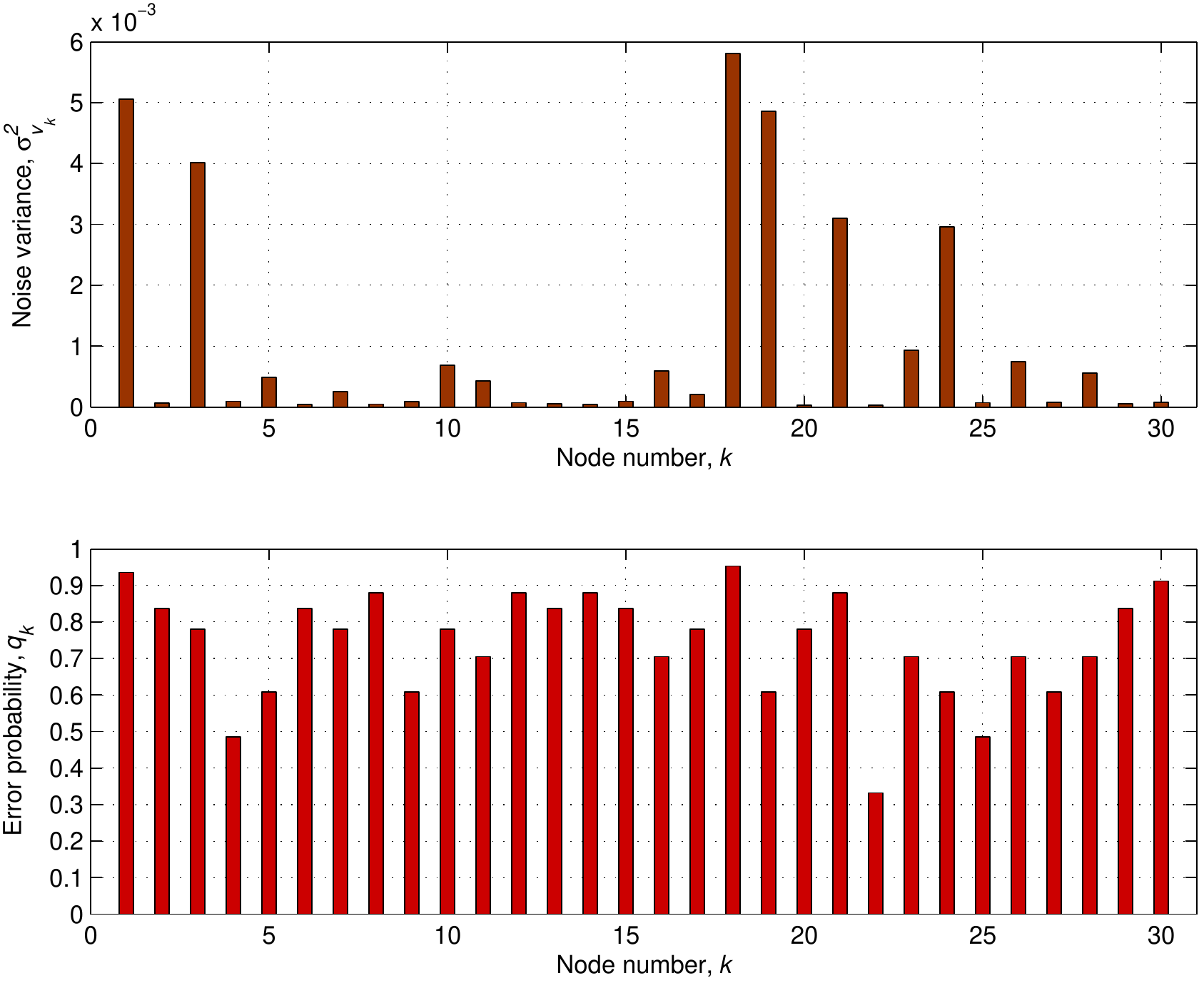}
\caption {Noise power profile $\sigma^2_{v_k}$ (top) and MAC-collision based error probability $p_k$ corresponding to backoff parameters $R=1$ and $CW=3$ for 30-node network in Fig.~\ref{topo_30_dense.pdf}.}
\label{error_profile.pdf}
\end{figure}
\begin{figure}[htbp]
\centering
\includegraphics[width=0.5\columnwidth] {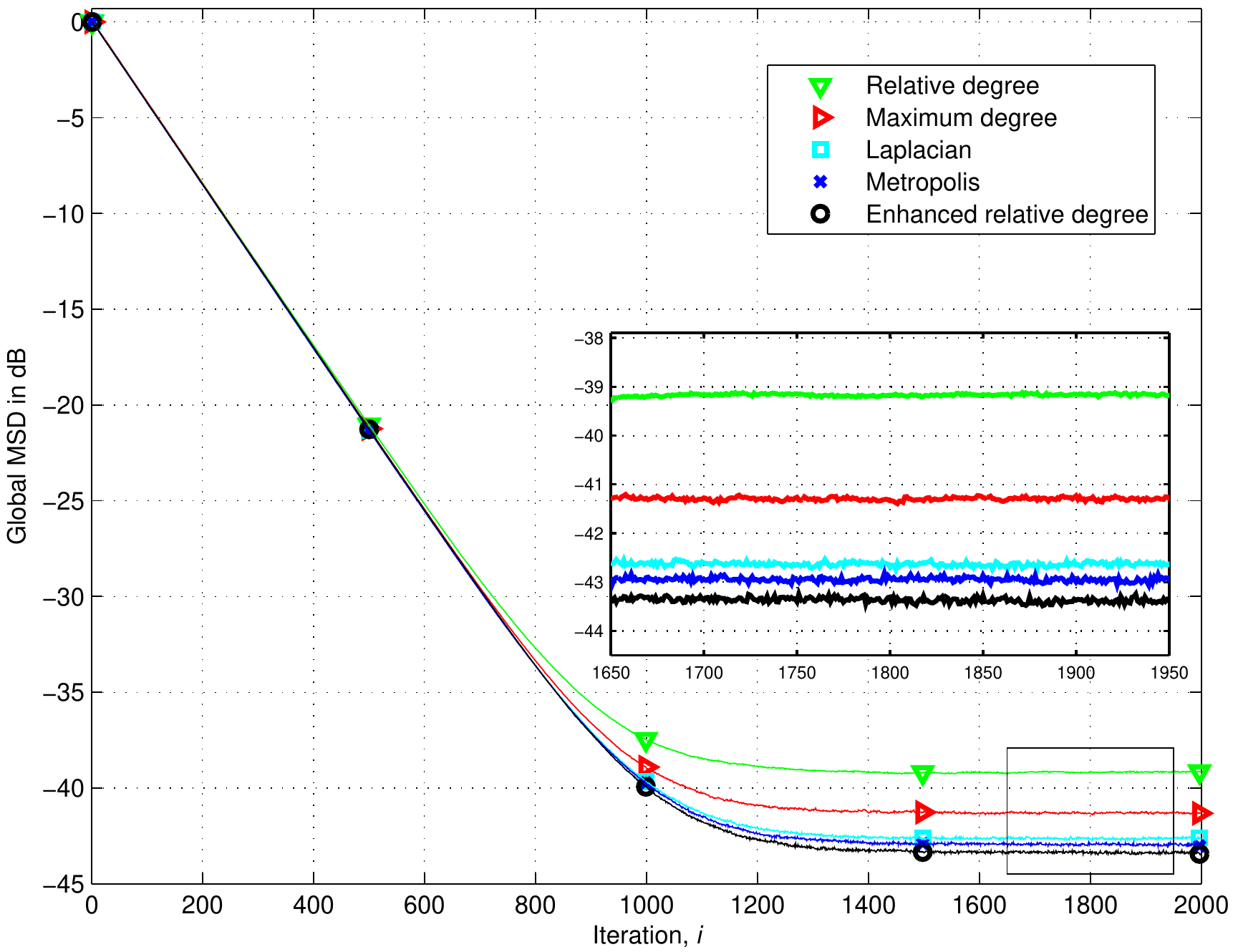}
\caption {Learning curve in terms of the global MSD in dB for different combining rules in 30-node network. The magnified image shows the performance improvement at the steady-state when using the enhanced relative degree combining rule.}
\label{learn_30.pdf}
\end{figure}
In the second phase of simulations, we concentrate on the MAC-level errors. We consider a medium-scale network including 30 nodes randomly placed in a square region with side $S=100$ units as shown in Fig.~\ref{topo_30_dense.pdf}. Nodes communicate with neighbors located within the range of $25$ units.
Initially, we perform a discrete-event simulation with $R=1$ and $CW=3$ to quantify error probabilities $q_k$ which is then verified using model (\ref{collision_link})-(\ref{loss probability}). Fig.~\ref{error_profile.pdf} (bottom) shows the profile of the error probability across the network. It is seen that nodes located in dense areas experience more errors than those in scattered regions. We now simulate the diffusion estimation algorithm for the corresponding MAC level errors shown at the bottom of Fig.~\ref{error_profile.pdf}. Noise levels are generated randomly from $[10^{-5},10^{-2}]$ and depicted in Fig.~\ref{error_profile.pdf} (top). We assume that the variances of input regressors are identical across nodes, i.e., $\sigma^2_u=\sigma^{2}_{u_k}=0.5$. We select $M=200$ and identical step-sizes: $\mu=\mu_k=0.01$.

We apply several combining policies that require only the degrees of nodes such as relative degree, Laplacian, Metropolis, maximum degree and enhanced relative degree. Learning curves in terms of the global MSD are shown in Fig.~(\ref{learn_30.pdf}). We observe that the diffusion estimation algorithm achieves the best performance in terms of the steady-state global MSD when using the enhanced relative degree rule.

\section{Conclusion}
\label{con}
We formulate the problem of distributed estimation based on the diffusion cooperation scheme over adaptive networks in the presence of transmission errors. We propose a theoretical framework and derive the closed-form expressions for the local and global steady-state MSD's under the assumption of imperfect information flow. Compared to the related work in the context of diffusion algorithms with error-free information exchange, the proposed analysis has less complexity and ensures scalability in terms of the input regressor size and the network size. Additionally, the present study does not impose the independence assumption between the observation vectors which in turn allows us to examine the performance measures of the distributed nodes with spatially correlated regressors. Simulation and analysis verify that a well-designed diffusion estimation algorithm will converge slower, achieving a higher steady-state MSD as a result of transmission errors. More importantly, we conclude that the local and global steady-state MSD curves are not necessarily monotonically increasing functions of the error probability. We also derive practically applicable sufficient conditions to assure the stability of diffusion LMS strategies with imperfect information sharing. Finally, we study a practical case scenario where errors occur at the MAC layer and introduce the enhanced relative degree to mitigate the negative effects of such errors.
 
\appendices
\section{Derivation of (\ref{ewss_f})}
\label{app_ma}

Define the one-sided $z\text{-transform}$ of the weight vector $\boldsymbol{w}_{k}$ as $\mathcal{W}_{k}=\mathcal{Z} \{E[\boldsymbol{w}_{k}]\}$. Taking the $z$-transform of (\ref{recursivewk}), we arrive at 
\begin{equation}
\label{zrecursivewk}
(1-a_{k,k} \rho_{k} z^{-1} ) \mathcal{W}_{k} - z^{-1}\sum_{  l \in \mathcal{N}, l \neq k } { a_{k,l} \mathcal{W}_{l} }=  \frac {  \boldsymbol{c}_{k}  } {  1-z^{-1}  }.
\end{equation} 
Thus, we can write the set of equations (\ref{setofequationswbar}) as follows:

\begin{equation}
\label{setofequationswbar}  
\underbrace{\begin{bmatrix}
  (1-z^{-1}a_{1,1}\rho_{1})\mathcal{I}_{M} & -z^{-1}a_{1,2}\rho_{1}\mathcal{I}_{M} & \cdots & -z^{-1}a_{1,N}\rho_{1}\mathcal{I}_{M} \\
  -z^{-1}a_{2,1}\rho_{2}\mathcal{I}_{M} & (1-z^{-1}a_{2,2}\rho_{2})\mathcal{I}_{M} & \cdots & -z^{-1}a_{2,N}\rho_{2}\mathcal{I}_{M} \\
  \vdots  & \vdots  & \ddots & \vdots  \\
  -z^{-1}a_{N,1}\rho_{N}\mathcal{I}_{M} & -z^{-1}a_{N,2}\rho_{N}\mathcal{I}_{M} & \cdots & (1-z^{-1}a_{N,N}\rho_{N})\mathcal{I}_{M}
 \end{bmatrix}}_{{\left[ \mathcal{E} \right]}_{NM \times NM}}
 \underbrace{\begin{bmatrix}
 \mathcal{W}_1\\
 \mathcal{W}_2\\
 \vdots \\
 \mathcal{W}_N
 \end{bmatrix}}_{\left[ \bar{\mathcal{W}} \right]_{NM \times 1}} = \underbrace{\begin{bmatrix}
 \frac{\boldsymbol{c}_{1}} {1-z^{-1}} \\
  \frac{\boldsymbol{c}_{2}} {1-z^{-1}} \\
  \vdots \\
   \frac{\boldsymbol{c}_{N}} {1-z^{-1}}
 \end{bmatrix}}_{\left[ \mathcal{F} \right]_{ NM \times 1}}.
\end{equation} For convenience, we define matrix $\mathcal{E}_{n,s}^{\prime}$ to collect $a_{k,l}~\text{and}~\rho_k,~k,l \in \mathcal{N}$ as $\mathcal{E}_{n,s}^\prime=[a_{k,l}\rho_{k}]_{N\times N}$. Using the Cramer's rule we obtain the following expression for $\mathcal{W}_k$
\begin{align}
\label{wbarkk}
\mathcal{W}_{k}=\frac{det(\mathcal{E}_{n,\mathcal{F}_{k}})} {det(\mathcal{E}_{n})} \boldsymbol{w}^{o},~k=1,2, \ldots, N,
\end{align} where
\begin{align}
\label{ensnew}
\mathcal{E}_{n}=\mathcal{I}_N - z^{-1}\mathcal{E}_{n,s}^\prime ,
\end{align} and $\mathcal{E}_{n,\mathcal{F}_i}$ is obtained by replacing the $i^{th}$ column of $\mathcal{E}_{n}$ by $\mathcal{F}_{n}$ defined as
\begin{equation}
\label{matrixfn}
\mathcal{F}_{n}=
\frac{1}{1-z^{-1}}
  \begin{bmatrix}
  \mu_{1} \sigma^{2}_{u_1}~\mu_{2} \sigma^{2}_{u_2}~\ldots~\mu_{N} \sigma^{2}_{u_N}
  \end{bmatrix}^T.
 \end{equation}  Denote $E[\boldsymbol{w}_{k,s}]$ as the expectation of the weight vector of the $k^{th}$ node at the steady-state. Then, we yield
\begin{equation}
\label{ewss}
E[\boldsymbol{w}_{k,s}]=\frac{det(\mathcal{E}_{n,s,\mathcal{F}_{k}})} {det(\mathcal{E}_{n,s})} \boldsymbol{w}^{o},~k=1,2, \ldots, N,
\end{equation} where $\mathcal{E}_{n,s}=\mathcal{I}_N -\mathcal{E}_{n,s}^\prime$ and $\mathcal{E}_{n,s,\mathcal{F}_{i}}$ is obtained by replacing the $i^{th}$ column of $\mathcal{E}_{n,s}$ by the $N \times 1$ column vector $\mathcal{F}_{n,s}$ whose $k^{th}$ element is $\mu_{k} \sigma^{2}_{u_k}$. Recall the following property of determinant: if any column of the determinant is replaced by a new column which is a linear combination of all columns, then the value of the determinant is not altered. As a result, for any arbitrary determinant and considering the sum of all columns as a particular linear combination of the columns, we can write
\begin{align}
\label{detp1}
\begin{vmatrix}
\mathcal{C}_1~\mathcal{C}_2~\cdots~\mathcal{C}_i~\cdots~\mathcal{C}_{Q}
 \end{vmatrix} = \begin{vmatrix}
\mathcal{C}_1~\mathcal{C}_2~\cdots~\mathcal{C}_i^{'}~\cdots~\mathcal{C}_{Q}
 \end{vmatrix},
\end{align} where $\mathcal{C}_i^{'}=\sum_{j=1}^{Q}{\mathcal{C}_{j}},$ for all $i=1,2,\ldots ,Q$.
Let us denote each column of $\mathcal{E}_{n,s}$ by $\mathcal{E}_{j}$ and find the sum of the columns of $\mathcal{E}_{n,s}$ as follows:
\begin{align}
\sum_{j \in \mathcal{N}}{\mathcal{E}_{j}}&=[ 1-\rho_{1}\sum_{l \in \mathcal{N}} {a_{1,l}}~\ldots~ 1-\rho_{N}\sum_{l \in \mathcal{N}} {a_{N,l}} ] ^{T} \nonumber \\
&=[ 1-\rho_{1} ~ \ldots ~  1-\rho_{N} ]^{T}=\mathcal{F}_{n,s}.
\end{align}
Notice that in the first and second step, we used (\ref{sumis1}) and (\ref{rhogeneral}) respectively. Hence, replacing any column of the $\mathcal{E}_{n,s}$ by $\mathcal{F}_{n,s}$ does not alter the value of the determinant of $\mathcal{E}_{n,s}$. The expressions in (\ref{ewss_f}) is then derived from (\ref{ewss}).

\section{Derivation of~(\ref{klgeneral_spatial_con})}
\label{app_ew4uw}

We begin by replacing (\ref{nodekphi})-(\ref{nodekd}) in the weight update rule (\ref{nodek}) which gives the following equation:
\begin{align}
\label{nodeksimple}
\boldsymbol{w}_{k,i+1}=~&a_{k,k,i}\boldsymbol{w}_{k,i}+ \sum \limits _{l \in \mathcal{N}_{k}\setminus\{k\}}{a_{k,l,i} \boldsymbol{w}_{l,i}}+\mu_{k} v_{k,i} \boldsymbol{u}_{k,i} + \mu_{k} \boldsymbol{u}_{k,i} \boldsymbol{u}_{k,i}^{T} \boldsymbol{w}^{o} \nonumber \\
& -a_{k,k,i} \mu_{k} \boldsymbol{u}_{k,i} \boldsymbol{u}_{k,i}^{T} \boldsymbol{w}_{k,i} - \sum \limits _{l \in \mathcal{N}_{k}\setminus\{k\}}{a_{k,l,i} \mu_{k} \boldsymbol{u}_{k,i} \boldsymbol{u}_{k,i}^{T} \boldsymbol{w}_{l,i}}.
\end{align}
Followed by multiplying recursion~(\ref{nodeksimple}) with correct indexes and taking the conditional expectation given that event $e_{j}$ occurs during data transmission, we arrive at~(\ref{klgeneral_spatial_con}).  

Note that in order to compute expressions of the form $E[\boldsymbol{w}_{m,i}^{T}\boldsymbol{u}_{l,i}\boldsymbol{u}_{l,i}^{T}\boldsymbol{u}_{h,i}\boldsymbol{u}_{h,i}^{T}\boldsymbol{w}_{n,i}]$, we use the independence assumption in the context of adaptive filters, i.e., the statistical correlations between the regressor vectors and the weight vectors is negligible \cite{haykin2002,sayed}. Also recall that the observation vectors are assumed to be temporally independent identically distributed (i.i.d.) white Gaussian random variables. This enables us to use the Gaussian moment factoring theorem \cite{papoulis1991}. With these in mind and omitting the time index $i$ for simplicity, a term of the form $E[\boldsymbol{w}_{m}^{T}\boldsymbol{u}_{l}\boldsymbol{u}_{l}^{T}\boldsymbol{u}_{h}\boldsymbol{u}_{h}^{T}\boldsymbol{w}_{n}]$ can be written as:
\begin{align}
\label{w u1 u1 u2 u2 w 1}
E[\boldsymbol{w}_{m}^{T}\boldsymbol{u}_{l}\boldsymbol{u}_{l}^{T}\boldsymbol{u}_{h}\boldsymbol{u}_{h}^{T}\boldsymbol{w}_{n}]&=E[\sum_{k}{w_{m,k}u_{l,k}} \sum_{j}{u_{l,j}u_{h,j}} \sum_{i}{u_{h,i}w_{n,i}}] \nonumber \\
& =\sum_{k,j,i}{E[w_{m,k}u_{l,k}u_{l,j}u_{h,j}u_{h,i}w_{n,i}]}\nonumber \\
& =\sum_{k,j,i}{E[w_{m,k}w_{n,i}] E[u_{l,k}u_{l,j}u_{h,j}u_{h,i}]},
\end{align} where the notations $w_{m,k},~k=1, \ldots , M$ and $u_{l,k},~k=1, \ldots , M$ are used to represent the $k^{th}$ elements of vectors $\boldsymbol{w}_{m}$ and $\boldsymbol{u}_{m}$ respectively. Notice that in the last step of Equation~(\ref{w u1 u1 u2 u2 w 1}), the independence assumption is used. Using the Gaussian moment factoring theorem, it can be verified that
\begin{align}
\label{w u1 u1 u2 u2 w 3}
E[\boldsymbol{w}_{m}^{T}\boldsymbol{u}_{l}\boldsymbol{u}_{l}^{T}\boldsymbol{u}_{h}\boldsymbol{u}_{h}^{T}\boldsymbol{w}_{n}]&=\sum_{i}{E[w_{m,i}w_{n,i}] (\sigma^2_{u_l} \sigma^2_{u_h} + 2 \sigma^{4}_{u_{lh}}) }+\sum_{i}{E[w_{m,i}w_{n,i}] (M-1) \sigma^{4}_{u_{lh}}}\nonumber \\
& =[\sigma^2_{u_l} \sigma^2_{u_h}+(M+1)\sigma^{4}_{u_{lh}} ] \sum_{i}{E[w_{m,i}w_{n,i}]} \nonumber \\
& =[\sigma^2_{u_l} \sigma^2_{u_h}+(M+1)\sigma^{4}_{u_{lh}} ] E[\boldsymbol{w}_{m}^{T} \boldsymbol{w}_{n}]. 
\end{align}

\section{Derivation of (\ref{wks_app_ms})}
\label{app_ms}

Using (\ref{ddwo}), we rewrite $\mathcal{D}_{w^{o}}$ as follows:
\begin{equation}
\label{dwoi}
[\mathcal{D}_{w^o}]_{i}=z^{-1}\sum_{m \in \mathcal{N}} {c_{j,om} \boldsymbol{w^{o}}^{T} \mathcal{W}_{m}} + \frac{\nu_{j} \boldsymbol{w^{o}}^{T} \boldsymbol{w^{o}} } {1-z^{-1}},~ i=1,\ldots ,Q,
\end{equation}  where index $j$ selects the proper coefficient for each index $i$ according to the permutation that is used to obtain the set of equations in (\ref{setofequ}). The key is to prove that 
\begin{equation}
\label{zdwoi}
\lim_{z \to 1} {(z-1)[\mathcal{D}_{w^o}]_{i}}= \boldsymbol{w^{o}}^{T} \boldsymbol{w^{o}} \sum_{m \in \mathcal{N}} {c_{j,om}+\nu_{j}},
\end{equation} for all $i=1,2, \ldots ,Q$. Denote by $\mathcal{E}_{n,i}$ the $i^{th}$ column of $\mathcal{E}_{n}$. Then it holds that $\mathcal{E}_{n}^{'}=\sum_{i \in \mathcal{N}} {\mathcal{E}_{n,i}}=[1-z^{-1}\rho_{j}]_{j}$.
Using $\rho_{j}=1-\mu_{j} \sigma^{2}_{u_{j}}$ and (\ref{matrixfn}), we obtain:
\begin{align}
\label{eprimen}
[\mathcal{E}_{n}^{'}]_{j}&=1-z^{-1}(1-\mu_{j}\sigma^{2}_{u_{j}}) \nonumber \\
&=1-z^{-1}(1-(1-z^{-1})[\mathcal{F}_{n}]_{j}) \nonumber \\
&=(1-z^{-1})(1+z^{-1}[\mathcal{F}_{n}]_{j}).
\end{align}
We know that 
\begin{align}
det(\mathcal{E}_{n})&=\begin{vmatrix} \mathcal{E}_{n,1}~\cdots~\mathcal{E}_{n}^{'}~\cdots~\mathcal{E}_{n,N}\end{vmatrix} \nonumber \\
&=(1-z^{-1}) \begin{vmatrix} \mathcal{E}_{n,1}~\cdots~ 1+z^{-1}\mathcal{F}_{n} ~\cdots~\mathcal{E}_{n,N}\end{vmatrix} \nonumber \\
&=(1-z^{-1})(\begin{vmatrix} \mathcal{E}_{n,1}~\cdots ~ 1 ~ \cdots~\mathcal{E}_{n,N}\end{vmatrix} + z^{-1}\begin{vmatrix} \mathcal{E}_{n,1}~\cdots ~ \mathcal{F}_{n} 
 ~ \cdots~\mathcal{E}_{n,N}\end{vmatrix}) \nonumber \\
 &=(1-z^{-1})(\begin{vmatrix} \mathcal{E}_{n,1}~\cdots ~ 1 ~ \cdots~\mathcal{E}_{n,N}\end{vmatrix}+z^{-1}\begin{vmatrix} \mathcal{E}_{n},\mathcal{F}_{n} \end{vmatrix}). \nonumber
\end{align}
Therefore (\ref{wbarkk}) can be written as follows:
\begin{align}
\label{wmzprove}
\mathcal{W}_{i}=\frac{\boldsymbol{w}^{o}} {(1-z^{-1}) \zeta(z) +z^{-1}(1-z^{-1})},
\end{align} where
\begin{align}
\label{zeta}
\zeta(z)=\frac {det(\mathcal{E}_{n,1}~\cdots ~ 1 ~ \cdots~\mathcal{E}_{n,N})}{det(\mathcal{E}_{n},\mathcal{F}_{i})}.
\end{align} With these, (\ref{dwoi}) becomes
\begin{align}
\label{mathcaldwopr}
[\mathcal{D}_{w^o}]_{i} = \frac {\boldsymbol{w^{o}}^{T} \boldsymbol{w^{o}}} {1-z^{-1}} [   z^{-1} \sum_{m \in \mathcal{N}} { { \frac {c_{j,om}} { \zeta(z)+z^{-1} } } } +\nu_{j}     ],
\end{align} and we get
\begin{align}
\label{limitmathcaldwopr}
\lim_{z \to 1} {(z-1) [\mathcal{D}_{w^o}]_{i} } =\lim_{z \to 1} { z \boldsymbol{w^{o}}^{T} \boldsymbol{w^{o}} [ z^{-1}  \sum_{m \in \mathcal{N}} { { \frac {c_{j,om}} {  \zeta(z) +z^{-1} } } } +\nu_{j} } ].  
\end{align}
Note that 
\begin{align}
\label{note}
det(\mathcal{E}_{n},\mathcal{F}_{i})=\frac{1}{1-z^{-1}}det(\mathcal{E}_{n,1}~\cdots ~ [\mu_{j} \sigma^{2}_{u_{j}}]_{i\in \mathcal{N}} ~ \cdots~\mathcal{E}_{n,N}). \nonumber
\end{align} Therefore, $\zeta(z)$ can be expressed as follows
\begin{align}
\zeta(z)=\frac{(1-z^{-1}) det(\mathcal{E}_{n,1}~\cdots ~ 1 ~ \cdots~\mathcal{E}_{n,N}) }  { det(\mathcal{E}_{n,1}~\cdots ~ [\mu_{i} \sigma^{2}_{u_{i}}]_{i\in \mathcal{N}} ~ \cdots~\mathcal{E}_{n,N}) },
\end{align} and we arrive at (\ref{zdwoi}) under one of the following conditions: either If we have that $[\mu_{i} \sigma^{2}_{u_{i}}]_{i\in \mathcal{N}}=\mu \sigma^{2}_{u}$ or we have that ${ det(\mathcal{E}_{n,1}~\cdots ~ [\mu_{i} \sigma^{2}_{u_{i}}]_{i\in \mathcal{N}} ~ \cdots~\mathcal{E}_{n,N}) } \neq 0 $. In order to proceed, we use (\ref{zdwoi}) to obtain the following expression for $c_{w^{o},s}$
\begin{equation}
\label{cwosprove}
c_{w^o,s}=\boldsymbol{{w}^o}^T \boldsymbol{w}^o \begin{vmatrix} \mathcal{C}_1 ~ \mathcal{C}_2 ~ \cdots ~ \mathcal{D}_s ~ \mathcal{C}_Q \end{vmatrix},
\end{equation} where $[\mathcal{D}_{s}]_i= \sum_{m \in \mathcal{N}} {c_{j,om}+\nu_{j}},~ i=1,\ldots ,Q$. After some algebra and using (\ref{coeff_general_N_nodes_klom}) and (\ref{coeff_general_N_nodes_kkom}) to replace $c_{kl,om},~k,l,m \in \mathcal{N}$ and then using $\sum_{l \in \mathcal{N}} {a_{k,l}}=1,~k=1,2,\ldots,N$ we obtain $[\mathcal{D}_{s}]_i=1-\eta_j,~i=1,\ldots ,Q$. Using (\ref{coeff_general_kl}), (\ref{coeff_general_klep}), (\ref{coeff_general_klnu}), (\ref{coeff_general_kk}) and (\ref{coeff_general_kknu}) we arrive at
\begin{align}
2\epsilon_{k}-\nu_{k}&=1-\eta_{k},\quad k=1,2,\ldots,N, \\
\epsilon_{k}+\epsilon_{l}-\nu_{kl}&=1-\eta_{kl},\quad k,l \in \mathcal{N}, k \neq l.
\end{align} With these, the sum of all columns of $\mathcal{I}_N-\mathcal{C}^\prime$ according to (\ref{cs}) becomes:
\begin{align}
\sum_{j=1}^{Q} {\mathcal{C}_{j}}&=[ 1-\sum_{k,l \in \mathcal{N}} {c_{11,kl}}~ \ldots~ 1-\sum_{k,l \in \mathcal{N}} {c_{N-1 ~ N,kl}}]^{T}. 
\end{align} Using (\ref{coeff_general_N_nodes}), (\ref{coeff_general_N_nodes_1}), (\ref{coeff_general_N_nodes_2}) and (\ref{coeff_general_N_nodes_3}), it can be checked that
\begin{align}
\sum_{j=1}^{Q} {\mathcal{C}_{j}} &=\bigg[ 1-\eta_{1} \sum_{j \in \mathcal{V}} {p_{j}( \sum_{m \in \mathcal{N}} {a_{1,m,j}})^{2}}~ \ldots ~\nonumber \\
&\qquad 1-\eta_{N-1~N} \sum_{j \in \mathcal{V}} {p_{j}( \sum_{m \in \mathcal{N}} {a_{N-1,m,j} \sum_{n \in \mathcal{N}} {a_{N,n,j}}    } )} \bigg]^{T} \nonumber \\
&=\bigg[
 1-\eta_{1}~\ldots~ 1-\eta_{N-1~N} 
 \bigg]^{T}=\mathcal{D}_{s}.
\end{align} Recalling the determinant property as stated in (\ref{detp1}), we conclude that replacing any of the columns of $\mathcal{I}_N-\mathcal{C}^\prime$ by $\mathcal{D}_s$ does not modify the value of its determinant, i.e., $|\mathcal{C}_{1}~\cdots~\mathcal{D}_{s}~\cdots~\mathcal{C}_Q|=|\mathcal{C}|$. As a result, equality (\ref{cwos}) holds.


\bibliographystyle{IEEEtran}
\bibliography{limit_13_single}



%
%
%

\end{document}